\documentclass[letterpaper]{article}
\pdfoutput=1
\usepackage{jheppub}
\usepackage[utf8]{inputenc}
\usepackage{amsmath,amssymb,mathrsfs}
\usepackage{graphicx}
\usepackage{array}
\usepackage{slashed}
\usepackage{appendix}
\usepackage{booktabs,multirow}
\usepackage{enumerate}
\usepackage{xcolor}
\usepackage{silence}
\usepackage{float}
\graphicspath{{../Mathematica/}}

\newcommand{\bea}{\begin{eqnarray}}
\newcommand{\eea}{\end{eqnarray}}
\newcommand{\kct}{\widetilde\kappa_3}
\newcommand{\kat}{\widetilde\kappa_1}
\newcommand{\phit}{\widetilde \phi}
\newcommand{\yt}{\widetilde y}
\newcommand{\Tt}{\widetilde T}
\newcommand{\Tca}{T_0}
\newcommand{\Tcat}{\widetilde T_0}
\newcommand{\Tcb}{T_\text{c}}
\newcommand{\Tcbt}{\widetilde T_\text{c}}
\newcommand{\Tst}{T_\star}
\newcommand{\HI}{H_\text{inf}}
\newcommand{\TRH}{T_\textsc{rh}}
\newcommand{\MP}{M_\text{pl}}
\newcommand{\NCMB}{N_\textsc{cmb}}
\newcommand{\dth}{\delta\theta}
\newcommand{\Ddth}{\Delta_{\delta\theta}}
\newcommand{\dIso}{\delta_\text{iso}}
\newcommand{\ODM}{\Omega_\textsc{dm}}

\title{Maximal axion misalignment from a minimal model}

\date{\today}
\author[a]{Junwu Huang,}
\author[a]{Amalia Madden,}
\author[a]{Davide Racco,}
\author[b]{and Mario Reig}

\affiliation[a]{Perimeter Institute for Theoretical Physics, 31 Caroline St.~N., Waterloo, Ontario N2L 2Y5, Canada}
\affiliation[b]{Instituto de Física Corpuscular (CSIC-Universitat de València),
C/ Catedrático José Beltrán 2, E-46980 Paterna (Valencia), Spain}
\emailAdd{jhuang@perimeterinstitute.ca}
\emailAdd{amadden@perimeterinstitute.ca}
\emailAdd{dracco@perimeterinstitute.ca}
\emailAdd{mario.reig@ific.uv.es}

\subheader{\normalsize
\vspace*{-2.em}
\begin{flushright}
IFIC/20-22
\end{flushright}
}

\abstract{
The QCD axion is one of the best motivated dark matter candidates. 
The misalignment mechanism is well known to produce an abundance of the QCD axion consistent with dark matter for an axion decay constant of order $10^{12}$ GeV. For a smaller decay constant, the QCD axion, with Peccei-Quinn symmetry broken during inflation, makes up only a fraction of dark matter unless the axion field starts oscillating very close to the top of its potential, in a scenario called ``large-misalignment''. 
In this scenario, QCD axion dark matter with a small axion decay constant is partially comprised of very dense structures.
We present a simple dynamical model realising the large-misalignment mechanism. 
During inflation, the axion classically rolls down its potential approaching its minimum. 
After inflation, the Universe reheats to a high temperature and a modulus (real scalar field) changes the sign of its minimum dynamically, which changes the sign of the mass of a vector-like fermion charged under QCD. 
As a result, the minimum of the axion potential during inflation becomes the {\it maximum} of the potential after the Universe has cooled through the QCD phase transition and the axion starts oscillating. 
In this model, we can produce QCD axion dark matter with a decay constant as low as $6\times 10^9\,{\rm GeV}$ and an axion mass up to 1 meV. 
We also summarise the phenomenological implications of this mechanism for dark matter experiments and colliders.
}

\preprint{\today}

\definecolor{AMcol}{rgb}{0.0, 0.42, 0.24}
\definecolor{JHcol}{rgb}{0.88, 0.14, 0.03}
\definecolor{DRcol}{rgb}{0.8, 0.04, 0.45}
\definecolor{MRcol}{rgb}{0.8, 0.56, 0.25}

\begin{document}
\maketitle

\section{Introduction and summary}
The QCD axion \cite{Peccei:1977hh,Wilczek:1977pj,Weinberg:1977ma,Dine:1981rt,Zhitnitsky:1980tq,Kim:1979if,Shifman:1979if} is one of the best motivated solutions to the strong CP problem of the Standard Model (SM).\footnote{For recent reviews on the QCD axion, see~\cite{diCortona:2015ldu,Irastorza:2018dyq,Hook:2018dlk,DiLuzio:2020wdo} and references therein.}
The strong CP problem originates from the experimental null measurement of the neutron electric dipole moment (EDM) \cite{Baker:2006ts,Afach:2015sja}.
This implies a strong constraint on the linear combination of CP violating angles in the SM, given by
\begin{equation}
\theta_{\rm SM} = \theta_{\rm QCD} + {\rm arg} \left[\det Y_u Y_d \right] \leq 10^{-10},
\end{equation}
where $\theta_{\rm QCD}$ is the QCD theta angle, and $Y_u$ and $Y_d$ are the up and down type Yukawa matrices respectively. 
The smallness of $\theta_{\rm SM}$ is particularly puzzling given the presence of another $\mathcal O(1)$ physical phase contained in the Yukawa matrices, namely the large complex phase of the Cabibbo-Kobayashi-Maskawa (CKM) matrix.

The axion solution to the strong CP problem consists of adding a new pseudoscalar particle $a$ coupling to the gluon field strength $G^{\mu\nu}$ as
\begin{equation}
\frac{a}{f_a} \frac{g_s^2}{32 \pi^2} G_{\mu\nu} \widetilde G^{\mu\nu},
\end{equation}
where $g_s$ is the strong coupling constant, $\widetilde G^{\mu\nu}=\tfrac 12 \varepsilon^{\mu\nu\rho\sigma} G_{\rho\sigma}$ is the dual field strength, and $f_a$ is the axion decay constant. 
The axion field dynamically sets the overall $\theta$ angle (combining the SM value and possible new physics contributions), and hence the neutron EDM, to zero regardless of the initial condition for $a$.

Various experiments have been proposed to search for the QCD axion through its couplings to nucleons \cite{Budker:2013hfa,Arvanitaki:2014dfa} and photons \cite{Wilczek:1987mv,Sikivie:1983ip,Dreyling-Eschweiler:2014mxa,TheMADMAXWorkingGroup:2016hpc,Andriamonje:2007ew,Anastassopoulos:2017ftl,Asztalos:2009yp,Armengaud:2014gea,Majorovits:2016yvk,Kahn:2016aff,Arvanitaki:2017nhi,Baryakhtar:2018doz,Chaudhuri:2018rqn} (see \cite{Graham:2015ouw,Arias:2012az,Irastorza:2018dyq} for a comprehensive review of current experimental efforts). 
Most of these experiments rely on the assumption that the axion particle is the dark matter (DM) in our Universe. 
Axion dark matter can be produced via cosmic strings if the Peccei-Quinn (PQ) symmetry breaking happens after inflation \cite{Davis:1986xc}, or through the misalignment mechanism if PQ symmetry is broken before inflation \cite{Preskill:1982cy,Abbott:1982af,Dine:1982ah}. 
These mechanisms can produce the current abundance of dark matter for an axion decay constant of order $f_a \gtrsim 10^{11}\, {\rm GeV}$, which implies $m_a \lesssim 100 \,{\rm \mu eV}$  \cite{Klaer:2017ond,Gorghetto:2018myk,Buschmann:2019icd} though recent studies suggest that the axion mass corresponding to the totality of dark matter may be larger than previously expected \cite{Gorghetto:2020qws}. 

There has been a recent surge in interest in probing axion dark matter for heavy axions with masses as large as $\mathcal{O} ({\rm eV})$~\cite{Arvanitaki:2017nhi,Baryakhtar:2018doz}, which calls for a different production mechanism if axions are to constitute the totality of DM.
Moreover, the last decade has seen an increasing interest in the formation of dense structures made of axions \cite{%
Kaup:1968zz,Ruffini:1969qy,Colpi:1986ye,%
Tkachev:1986tr,Hogan:1988mp,Kolb:1993zz,Kolb:1993hw,
Barranco:2010ib,
Eby:2014fya,
Braaten:2015eeu,
Chavanis:2016dab,
Visinelli:2017ooc,
Hertzberg:2018lmt%
} 
(see \cite{Eby:2019ntd} for a compact review on the topic of axion stars). 
The most elegant way to produce heavy axion DM while simultaneously creating such structures of axions is the `large-misalignment mechanism' \cite{Arvanitaki:2019rax}, where the axion field is very close to the top of the (approximate) cosine potential at $|\theta_i| \equiv a_i/f_a \simeq \pi$ when it starts oscillating.%
\footnote{We adopt the convention that $\theta$ ranges between $-\pi$ and $\pi$.}
This allows an enhancement of the axion density by delaying the oscillations of the axion field in the early universe, and an amplification of density perturbations through parametric resonance when the axion starts to oscillate during radiation domination, leading to much denser and more numerous small halos (see \cite{Arvanitaki:2019rax} for more details). 

The main drawback of the `large-misalignment mechanism' is the reappearance of a strong dependence on initial conditions. 
Such a strong dependence may potentially be justified with anthropic arguments \cite{Arvanitaki:2009fg,Arvanitaki:2019rax}. 
In this paper, we will present two simple models where the initial condition of $|\theta_i| \simeq \pi$ is set {\it dynamically} in the early Universe.
There have been a few recent proposals where this initial condition is set dynamically, relying on large sectors of physics beyond the SM \cite{Co:2018mho} or particular parameter choices and initial conditions \cite{Takahashi:2019pqf}. 
Our goal is to discuss a scenario that includes only the minimal ingredients required to set $|\theta_i| \simeq \pi$.

\begin{figure}[h!] \centering%
{Field $\phi$ settling in \textbf{true} vacuum (point \includegraphics[height=1em]{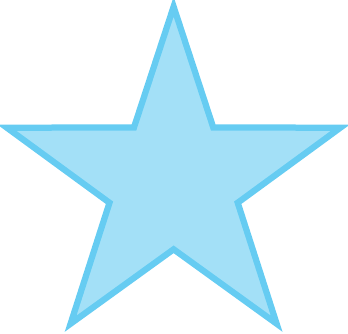} in Fig.~\ref{fig:MoneyPlot})} \smallskip\\%
\includegraphics[width=.8\textwidth]{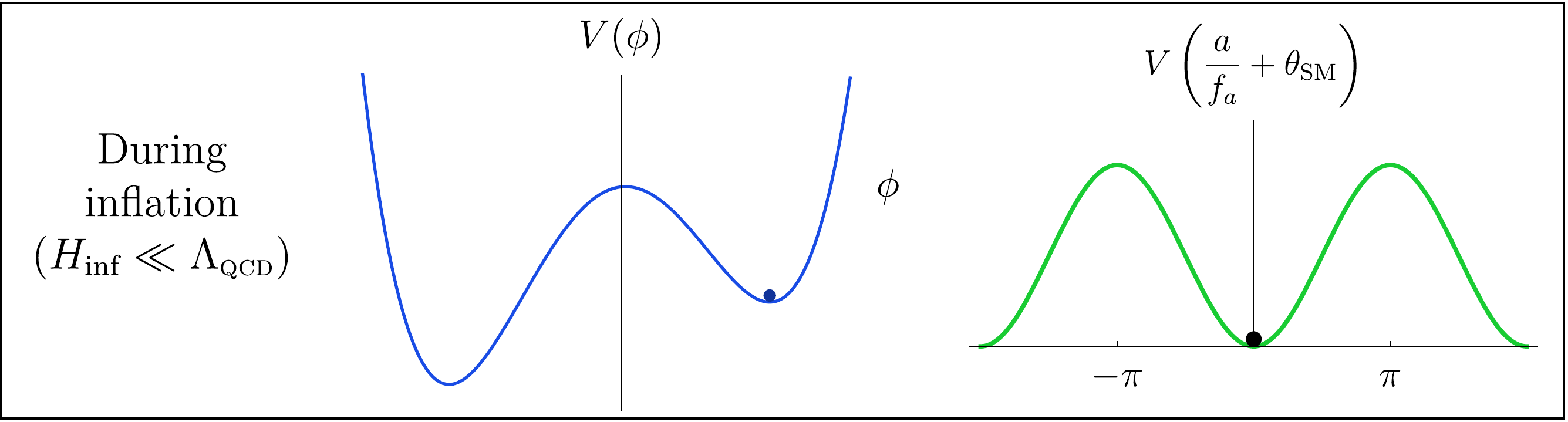}\smallskip\\%
\includegraphics[width=.8\textwidth]{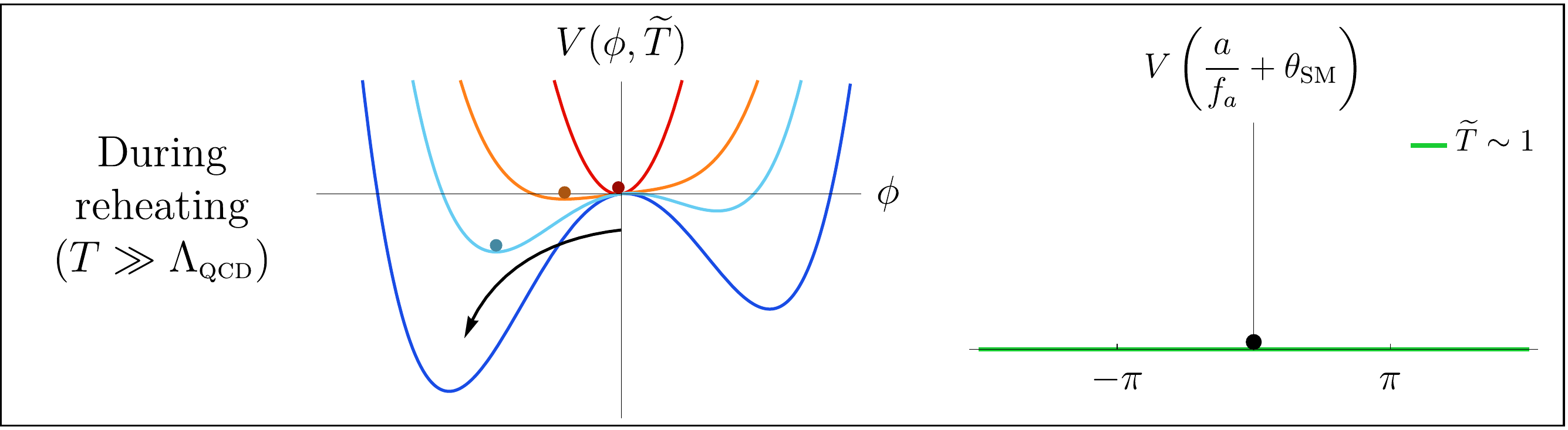} \smallskip\\%
\includegraphics[width=.8\textwidth]{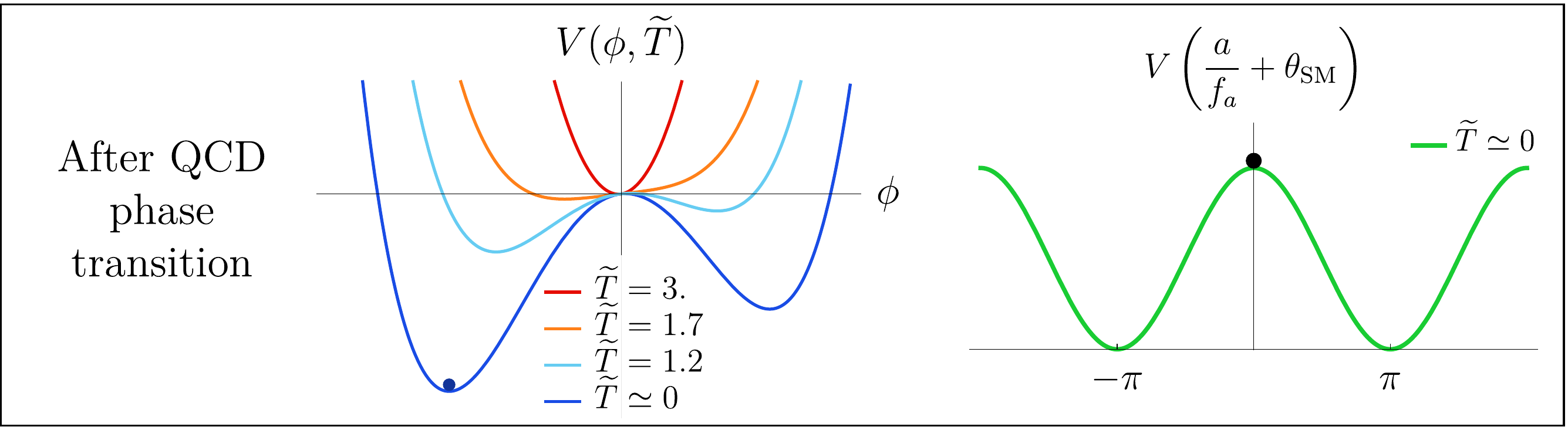} \smallskip\\%
\caption{Dynamics for the real scalar field $\phi$ and the axion field $a$ in the early Universe. 
Coloured lines show their potentials at specific temperatures, and the corresponding dots represent the field values at those times.
We show here the case in which the field $\phi$ lies in its false vacuum during inflation and ends up in its true vacuum today.
For visualisation purposes, the Yukawa coupling $y$ is taken to be positive in these figures.
\textbf{Top panel}: We assume that the Hubble rate during inflation is below the QCD scale, so that the QCD potential for the axion is turned on and $a$ lies in the minimum where the $x$-axis coordinate of the right-hand side plots $a/f_a+\theta_{\rm SM}=0$ (see Eq.~\eqref{theta}).
\textbf{Central panel}: During the thermal phase, the scalar $\phi$ and a coloured fermion $q$ are produced in the thermal bath and induce corrections to the effective potential $V(\phi)$ that lift the two minima of $\phi$. 
As the temperature decreases and thermal corrections diminish, for this choice of parameters of $V(\phi)$ (corresponding to the blue star in Fig.~\ref{fig:MoneyPlot}) the field $\phi$ settles in the true vacuum.
During this period $T\gg \Lambda_\text{QCD}$ and the axion potential is turned off.
\textbf{Bottom panel}: The field $\phi$ settles in its true vacuum, which has opposite sign with respect to its VEV during inflation. 
Accordingly, the mass of $q$ changes sign and this introduces a shift of $\pi$ in the argument of the mass matrix of coloured fermions. 
Therefore the axion potential shifts by $\pi$, whereas the  initial value of $a$ has not changed since inflation. 
The initial misalignment of the axion field is close to maximal. 
Note that in this panel, the mass of the fermion $q$ is negative, and the overall $\theta$ angle that we measure in EDM experiments, that is, $\theta = a/f_a+\theta_{\rm SM} + \arg \left[m_q\right]$, has a minimum at $\theta = 0$.
}%
\label{fig: phi true}%
\end{figure}

The requirement for setting any initial value of $\theta_i$ before the time of the QCD phase transition is a period in the early universe when the axion potential $V (a)$ is turned on. 
Assuming a standard cosmological history of inflation followed by a radiation dominated era, such a period can only occur during inflation.
At that time, the effective temperature (the Hubble scale during inflation $\HI$) of the SM sector can be much smaller than the QCD confinement scale $\Lambda_{\rm QCD}$:
\begin{equation}
\label{eq: H < LQCD}
\HI \ll \Lambda_{\rm QCD}\,.
\end{equation}
During this period, the potential $V(a)$ of the QCD axion field $a$ is turned on and the axion evolves with the equation of motion
\begin{equation}
\ddot{a} + 3 H \dot{a} + V'(a) = \eta(t) \,,
\end{equation}
where by $\eta(t)$ we denote the stochastic noise of $a$ due to quantum fluctuations during inflation, which imprint quantum jumps in $a$ of order $\HI/(2\pi)$ on time scales of order $\HI^{-1}$.
The axion can roll towards the minimum of its potential if the classical evolution over a Hubble time is larger than quantum fluctuations:
\begin{equation}
\HI^3 \ll V'(a) \sim m_a^2 f_a \,.
\end{equation}
As a result of Eq.~\eqref{eq: H < LQCD}, the QCD axion potential is turned on, and the axion can slowly roll down its potential towards its minimum during inflation. 
The approximate number of $e$-folds required to the axion to reach its initial position can be estimated as $\HI^2/m_a^2$, and in the region of interest of our parameter space turns out be of order $\gtrsim \mathcal O(10^3)$.
This classical evolution proceeds until quantum spreading of the wave function dominates over rolling, when the axion field is displaced from the minimum of its potential by an amount%
\footnote{The final equilibrium distribution, after a long time of inflationary evolution, yields a much smaller misalignment from the minimum of the potential during inflation $\delta \theta \sim \HI^2/(m_a f_a)$ \cite{Linde:2005ht,Graham:2018jyp,Guth:2018hsa}.}
\begin{equation}\label{initial_misalignment}
\dth \equiv \pi-|\theta_i| \simeq \frac{\HI^3}{m_a^2 f_a}\simeq 10^{-13}\left(\frac{\HI}{1\text{ eV}}\right)^3\left(\frac{f_a}{10^{10}\text{ GeV}}\right) \,.
\end{equation}
In order to set the initial condition $|\theta_i| \simeq \pi$, we need to flip the QCD axion potential between inflation and today, as can be seen in the right-hand panels of Fig.~\ref{fig: phi true}. 
This can be achieved by changing the sign of the masses of fermions charged under $SU(3)_{\rm C}$ (in a similar spirit to \cite{Co:2018mho}). 
For example, in the model presented here, we add to the SM a vector-like Dirac fermion $q$ with charge $(\mathbf 3,\mathbf 1,+ \tfrac 23)$ or $(\mathbf 3, \mathbf 1,-\tfrac 13)$ under the SM gauge group $SU(3)_{\rm C} \times SU(2)_{L}\times U(1)_{Y}$.
The Lagrangian terms that we add to the SM are
\begin{equation}
\begin{gathered}
\mathcal L \supset \mathcal L_\textsc{sm} + \mathcal L_{q} + V(\phi) \,, \\
\mathcal L_q \supset y \phi \, \overline q q + \text{(SM-$q$ mixing terms)},
\label{eq: Yukawa phi q}
\end{gathered}
\end{equation}
where $V(\phi)$ is the potential of a real scalar field $\phi$ (e.g.~a modulus for the fermion $q$), and the mixing terms of $q$ with the SM will be explicitly spelled in Eq.~\eqref{Yuk_Lag}. 
The Yukawa coupling $y$ is the only spurion that breaks chiral symmetry for the new fermion $q$, and the reference point $\phi=0$ is chosen to be where $q$ is massless.
The Lagrangian $\mathcal L_q$ has a $\mathbb Z_2$ symmetry under which the real scalar $\phi\leftrightarrow -\phi$ and the left-handed component of the Dirac fermion $q_L \leftrightarrow -q_L$, whereas the right-handed component of the Dirac fermion $q_R$ as well as all SM fields are singlets.
This $\mathbb Z_2$ symmetry is explicitly broken by the scalar potential $V(\phi)$, and is further spontaneously broken by $\langle\phi\rangle$.
With the inclusion of this pair of heavy quarks, the axion dynamically sets the linear combination
\begin{equation}\label{theta}
\theta = \frac{a}{f_a}+ \theta_{\rm SM} + {\rm arg} \left[m_q\right] 
\end{equation}
to zero, where $m_q = y \langle \phi \rangle$ is the mass of the heavy vector pair of quarks. 
As shown in the right panel of Fig.~\ref{fig: phi true}, if the sign of the vacuum expectation value (VEV) of the real scalar field $\phi$, and hence the mass of the heavy quark $q$, changes between the time of inflation and today, then the quantity ${\rm arg} \left[m_q\right]$ changes by $\pi$. 
This induces a shift of $\pi$ between the minimum $\theta_{i}$ of the axion potential during inflation and its minimum today, ensuring that the axion field starts oscillating from near the top of its potential when the QCD axion potential reappears at later times in the cosmological history. 
For a more detailed account of the cosmological evolution of the fields $\phi$ and $a$, see the captions of Figs.~\ref{fig: phi true}, \ref{fig: phi false} and \ref{fig: phi decayed}.\\
The modulus field $\phi$ can in principle have a rather complicated potential $V(\phi)$ with many minima, each corresponding to a different mass $m_q$ of the quark $q$.
In the following, we will consider a simple example of a potential $V (\phi)$, which has two minima on opposite sides with respect to $\phi=0$ and an \textit{explicitly} broken $\mathbb Z_2$ symmetry ($\phi \rightarrow -\phi$), as shown in the top panel of Fig.~\ref{fig: phi true}. 
The explicit breaking ensures the sign of $m_q$ to be physical.
The scalar potential at high temperature, however, has only a single minimum around $\phi = 0$. 
Therefore, as long as the Universe reheats to a large enough temperature, the high-temperature phase essentially erases all information about the sign of $m_q$ apart from the initial value of the axion field set during inflation. 
This can be realised if the maximum temperature achieved during reheating, denoted by $\TRH$, satisfies
\begin{equation}\label{RH_temp}
\TRH=\left(\frac{90}{\pi^2 g_\star}\right)^{1/4}\sqrt{\HI M_{\text{pl}}}\sim 30\text{ TeV}\left(\frac{\HI}{1\text{ eV}}\right)^{1/2}\gg \langle \phi \rangle \,,
\end{equation}
where we have assumed instantaneous reheating.

To summarise, as we will show in more detail in the next sections, our mechanism only requires the two following conditions to be satisfied:
\begin{enumerate}
\item The Lagrangian contains \textit{two} sources of explicit $\mathbb Z_2$ symmetry breaking.
\item This $\mathbb Z_2$ symmetry is approximately restored during radiation domination. 
\end{enumerate}
Once the two conditions are satisfied, we find that regardless of whether the Universe prefers to live in the lower energy (true) or the higher energy (false) minimum of the modulus potential $V(\phi)$ during inflation, that is, regardless of the \textit{measure} \cite{Linde:2006nw,Susskind:2012xf}, the thermal phase after reheating can make sure that the axion field starts oscillating from the top of its potential.
The measure problem, practically, pertains to the inability to compute the statistical probability to live in each individual minimum in an eternally inflating universe. 
In the simplest axion models~\cite{Dine:1981rt,Zhitnitsky:1980tq,Kim:1979if,Shifman:1979if}, the $\theta$ angle can acquire an initial value anywhere between $0$ and $2\pi$ during inflation, and a probability distribution for the parameter $\theta$ should be assigned over a continuous range. 
With the addition of the real scalar field $\phi$ and the vector-like quark $q$, the possible values for $\theta$ reduce from a continuous range to a discrete set.
The limitation of the choice to $\theta_i =0$ and $|\theta_i| =\pi$ might seem comforting, but does not qualitatively improve the situation, because we are still unable to determine whether there is a non-negligible probability of living in a given minimum. 
The addition of a single source of $\mathbb Z_2$ symmetry breaking in the potential of $\phi$ makes the post-inflationary evolution {\it deterministic}, and the Universe always settles in the true minimum of the zero-temperature potential of the modulus $\phi$. 
However, without specifying a measure, it is unclear whether this {\it deterministic} evolution is able to switch the vacuum of $\phi$ between inflation and today, and thereby whether the axion starts respectively from $|\theta_i| =\pi$ or $\theta_i =0$. 
The presence of two sources of $\mathbb Z_2$ symmetry breaking solves this problem, allowing the Universe to evolve into the minimum corresponding to either the higher energy or the lower energy minimum of the modulus potential during inflation. 
Once a measure is specified, prescribing whether $\phi$ during inflation lived in its higher or lower energy minimum, we can provide the regions of parameter space in our model for which $\phi$ ends up with probability one in a vacuum with opposite sign with respect to the one during inflation%
\footnote{Throughout this paper, we compute the probability for the axion to start oscillating in the late universe from close to its maximum, and we treat different choices of measure as the preference for $\phi$ to be in either the higher energy or the lower energy minimum during inflation. 
The assumption that a given measure would select with probability close to unity only one of the two minima is well motivated when their energy difference is much larger than the Hubble scale during inflation, but much smaller than the energy scale of inflation. 
We leave a more in-depth discussion to Sec.~\ref{sec: remarks}.}.
This ensures that the QCD axion can always begin its oscillations near $|\theta_i| =\pi$ after the QCD phase transition.
The evolution of $\phi$ and the axion $a$ for the model presented in Sec.~\ref{sec: model} is illustrated for the cases in which the measure prescribes $\phi$ to live in its higher (respectively lower) energy minimum during inflation in Fig.~\ref{fig: phi true} (respectively Figs.~\ref{fig: phi false} and \ref{fig: phi decayed}). 
In all cases, the result is that $|\theta_i| =\pi$.

\begin{figure}[t] \centering
\includegraphics[width=.9\textwidth]{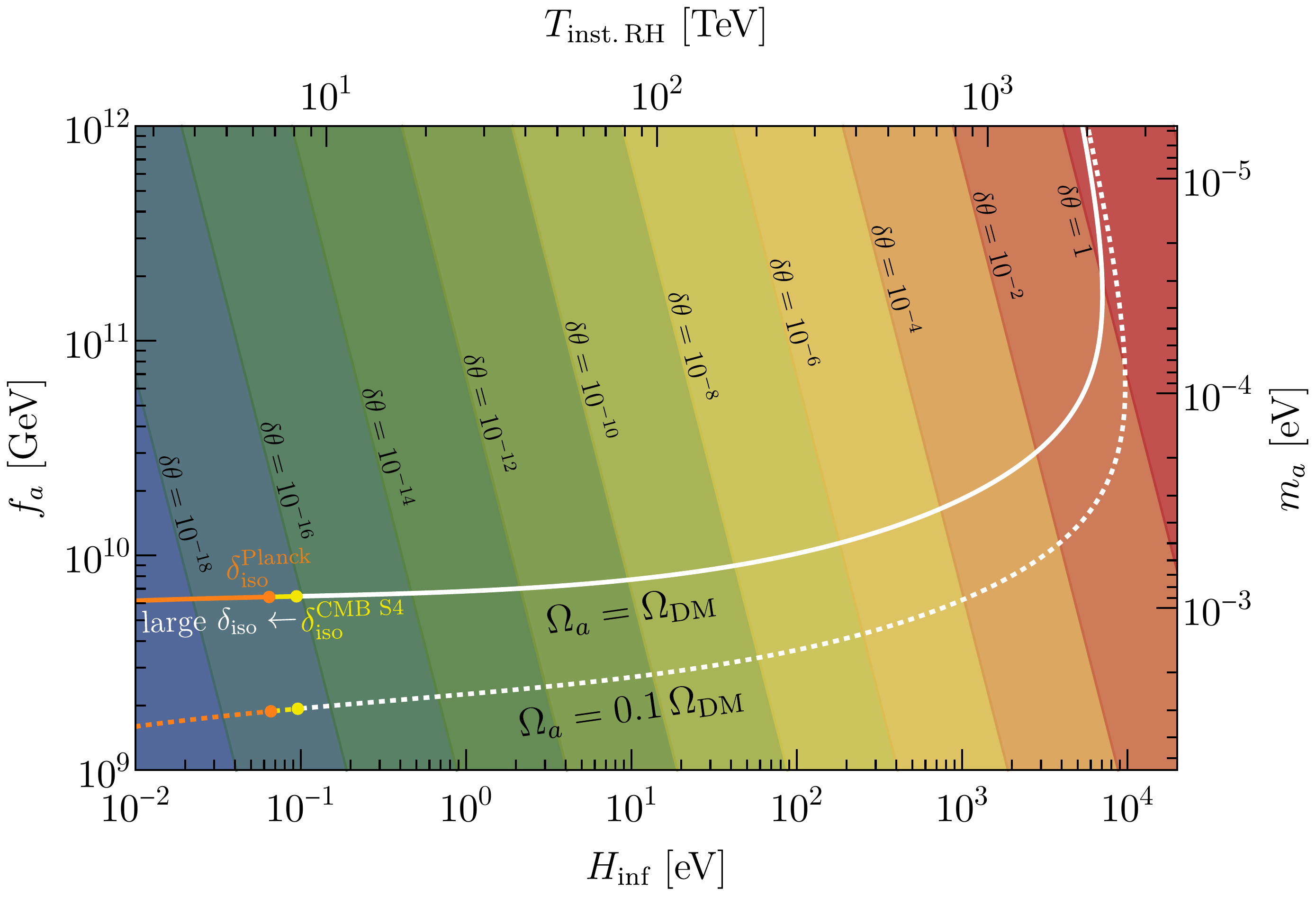}
\caption{Parameter space leading to an axion abundance $\Omega_a$ equal to the observed dark matter abundance $\Omega_\text{dm}$ (solid white line). 
For reference, we also show the line for $\Omega_a=0.1\Omega_\text{DM}$ (white dashed).
The axion abundance in the misalignment mechanism is set by $f_a$ and the displacement $\delta \theta$ of the axion field from the top of its potential as it starts oscillating. 
Contours of $\delta \theta$ are shown with coloured lines.
In our setup, where the typical size of $\delta \theta$ is set by the spread of the axion field around its minimum during inflation, we trade $\delta \theta$ for the Hubble rate during inflation $\HI$ (shown on the lower horizontal axis). 
For reference, we show on the upper horizontal axis the temperature implied by a given $\HI$ for instantaneous reheating.
Along the white lines, we colour in orange (respectively yellow) the parameter space excluded by Planck \cite{Akrami:2018odb} (respectively testable by CMB S4 \cite{Abazajian:2019eic}) searches of isocurvature fluctuations (figure adapted from \cite{Arvanitaki:2019rax}).}
\label{fig:AxionDM-abundance}
\end{figure}

In summary, we describe a simple model (introducing just a real scalar field $\phi$ and a pair of coloured heavy quarks $q$) that offers a \textit{dynamical} origin of an axion misalignment angle close to $\pi$.
After accounting for cosmological and experimental constraints, the two new particles contained in this model are expected to lie in a quite limited window between 1 TeV and $\mathcal O(10^2)$ TeV, not far from the energy range of future colliders.
As shown in Fig.~\ref{fig:AxionDM-abundance}, the QCD axion mass in this model would be up to a few meV, with an axion decay constant $f_a\lesssim 10^{10}$ GeV.
One of the key observational signatures of this scenario is associated with the presence of dense structures of axions in the DM distribution, which can be probed experimentally through the wide variety of techniques illustrated in \cite{Arvanitaki:2019rax}.

The paper is organised as follows. 
Sec.~\ref{sec: toy model} contains a toy example illustrating how the modulus $\phi$ can change its minimum in a simple theory with two sources of $\mathbb Z_2$ breaking. 
In Sec.~\ref{sec: model}, we discuss in detail the evolution of the scalar field $\phi$ with a generic quartic potential. 
Sec.~\ref{sec: pheno} summarises the phenomenology of the additional particles around the TeV scale. 
In Sec.~\ref{sec: remarks}, we comment on inflationary model building which {\it circumvents} the measure problem.
We include in the Appendices \ref{app: V finite T}, \ref{app: 3TC1} and \ref{app: vacuum decay} further details about the derivation of finite-temperature effects and the rate for vacuum decay.

\section{A toy example}
\label{sec: toy model}

In this section, we discuss a simple toy model where the mass of a fermion charged under $SU(3)_\text{C}$ can change sign. 
As we alluded to in the Introduction, in order to {\it deterministically} land either into the true or the false minimum, we require at least two independent sources of $\mathbb Z_2$ symmetry breaking in the Lagrangian of the real scalar field $\phi$. 
In this section, we discuss the simplest scenario where this can be achieved. 
The potential of the real scalar field $\phi$ is
\begin{equation}
\label{eq:ToyModel}
V(\phi) = \kappa_1^3\phi -\frac{\mu^2}{2}\phi^2 + \frac{\lambda}{4!}\phi^4 + g \phi  R,
\end{equation}
where $\kappa_1$ and $g$ are real, dimension one parameters that softly break the $\mathbb Z_2$ symmetry, $R$ is the Ricci scalar, and $\mu^2>0$. 
In this simple model, the sign and size of the parameter $g$ uniquely determine the field location preferred by the Universe during inflation, whereas the sign and size of the parameter $\kappa_1$ determine the evolution of the scalar field $\phi$ after the Universe thermalises. 
In principle, the scalar potential $V(\phi)$ could also contain a cubic term, given that the location $\phi = 0$ is defined as the field value where the fermion $q$ coupled to $\phi$ through a Yukawa coupling as in Eq.~\eqref{eq: Yukawa phi q} becomes massless. 
Such a cubic term, however, is generated with very small values if the tadpole $\kappa_1$ and the coupling $g$ are the only spurions of $\mathbb Z_2$ symmetry breaking.

During inflation, the relative difference in size of the vacuum energy of the two minima is determined by the non-minimal coupling between the scalar $\phi$ and the curvature $R$ (for small values of $\kappa_1$). 
The Universe lies in just one of these two minima, selected by the solution to the measure problem \cite{Linde:2006nw,Susskind:2012xf}. 
During the thermal phase following inflation, all information about which minimum the scalar field $\phi$ lives in during inflation is totally removed due to the thermal corrections to $V(\phi)$.
After inflation, the Universe lands in a minimum deterministically set by the sign of $\kappa_1$, which is the only significant breaking to the $\mathbb Z_2$ symmetry during radiation domination. 
The sign of $\kappa_1$ determines the location of the true minimum after inflation, which is also the minimum into which the Universe evolves as it cools down.

We acknowledge that we do not have a solution to the measure problem and, in particular, that we do not know whether the scalar field prefers to live in its true or false vacuum during inflation. 
However, for any given choice of measure, corresponding to a choice of the vacuum of $\phi$ during inflation (determined by the sign of $g$), our mechanism is always able to direct $\phi$ to a minimum of the opposite sign (selected through the choice of the sign of $\kappa_1$).%
\footnote{In principle, a single source of $\mathbb Z_2$ breaking $\kappa_1$ can lead to a change of minimum between the period of inflation and today if the solution to the measure problems prescribes the field to live in the false vacuum during inflation. 
This speaks of the high probability of the initial condition $\theta_i=\pi$ in some of the simplest landscapes.
In this paper, we offer a few deterministic scenarios with technically natural parameters.}
Since the signs of $g$ and $\kappa_1$ are totally independent, we can always choose their relative sign in a way that accommodates for the vacuum preferred during inflation according to the solution of the measure problem, and simultaneously make $\langle \phi\rangle$ change sign between inflation and today. 
We elaborate on this new type of model building which evades the measure problem in more detail in Sec.~\ref{sec: remarks}.

The necessary conditions to realise this mechanism are the following.
In order for the $\kappa_1^3 \phi$ term to contribute less to the potential energy than the term $g \phi R = -12 g\phi \HI^2$ during inflation, so that $\text{sign}[g]$ determines the vacuum of $\phi$ during inflation, we require
\begin{equation}
|\kappa_1| \lesssim \left(12 |g| \HI^2\right)^{1/3} \approx 1 \,{\rm GeV}\left(\frac{\HI}{1\,{\rm eV}}\right)^{2/3} \left(\frac{|g|}{0.1 \MP}\right)^{1/3} \,.
\label{eq: kappa1 inflation}
\end{equation}
On the other hand, in order for the Universe to be in the true vacuum {\it deterministically} during radiation domination before the axion starts to oscillate around the QCD phase transition, we impose that any region of false vacuum disappears before the axion start oscillating.
This condition can be estimated by imposing that the time scale for the collapse of the bubble is smaller than a Hubble time $H_\text{osc}^{-1}$ when the axion starts oscillating.
The time scale for the collapse of the bubble is given by the ratio of the bubble wall surface tension $\sigma_\text{wall}$ (of order $\sqrt \lambda v^3$, where $v=\langle \phi \rangle$ and we take for simplicity $\lambda \sim 1$) and the energy difference $\Delta V \sim \kappa_1^3 v$.
The condition $\sigma_\text{wall}/\Delta V \lesssim H_\text{osc}^{-1}$ implies a lower bound on $|\kappa_1|$:
\begin{equation}
|\kappa_1| \gtrsim \left( v^2 H_{\rm osc}\right)^{1/3} \approx 100 \,{\rm keV} \left(\frac{v}{10 \,{\rm TeV}}\right)^{2/3} \left(\frac{H_{\rm osc}}{10^{-9} \,{\rm eV}}\right)^{1/3} \,.
\label{eq: kappa1 thermal}
\end{equation}
For $|\kappa_1|$ in the range identified by Eqs.~\eqref{eq: kappa1 inflation} and \eqref{eq: kappa1 thermal}, the sign of the coupling $g$ determines the location of the true minimum during inflation, while the sign of the coupling $\kappa_1$ determines the true minimum that the Universe evolves to after reheating. 

Finally, we need to discuss whether this picture with a sizeable coupling to the Ricci scalar, a smaller linear term and a negligible cubic term is stable under radiative corrections.
The $\mathbb Z_2$ breaking coupling $g \phi  R \supset g\,\phi\, \partial h \partial  h /\MP^2$ generates linear and cubic terms in $V(\phi)$ through a loop of the graviton $h$.
By dimensional analysis, the size of the induced coefficients can be estimated as
\begin{equation}
\kappa_1 \simeq \left(\frac{g \Lambda^4}{16 \pi^2 \MP^2} \right)^{1/3}, \quad 
\kappa_3 \simeq \frac{g^3 \Lambda^4}{16 \pi^2 \MP^6},
\end{equation}
where $\Lambda$ is the cutoff of the theory. 
As a result, for a theory with cutoff $\Lambda \ll \MP$, the induced $\kappa_3$ is much less important than $\kappa_1$. 
The requirement that the radiatively induced linear term $\kappa_1^3 \phi$ is less relevant during inflation than the term $g\phi R = -12 g \phi \HI^2$ gives a lower bound on $\HI$ as a function of $\Lambda$:
\begin{equation}
12 |g| \HI ^2 \gtrsim \lvert \kappa_1^3\rvert \quad \leftrightarrow \quad 
\HI \gtrsim\frac{\Lambda^2}{8\pi \sqrt 3 \MP} \,.
\end{equation}
This condition allows $\Lambda$ to be raised up to a value slightly larger than the energy scale during inflation (independently of the value of $g$), resulting in a natural choice of scales.

In summary, it is possible to flip the sign of the VEV of $\phi$ between inflation and the late universe by introducing a coupling $g\phi R$ and a small linear term $\kappa_1^3 \phi$, independently of the solution to the measure problem.
The field $\phi$ always lands in its true vacuum (determined by the sign of $\kappa_1$) and the sign of its VEV during inflation can always be opposite to the one of the true vacuum by choosing a large enough coupling $g\phi R$ with a suitable sign.
For this mechanism to work, it is enough to require the conditions \eqref{eq: kappa1 inflation} and \eqref{eq: kappa1 thermal} on $\kappa_1$, and radiative corrections do not alter this picture.

We want to emphasize that, although in this toy example the potential of $\phi$ is not the same during inflation and today, this is not a necessary ingredient for our mechanism to work. 
As will become more apparent from the model presented in Sec.~\ref{sec: model}, our mechanism only requires the two following conditions.
\begin{enumerate}
\item There are two sources of $\mathbb Z_2$ symmetry breaking in the Lagrangian.
\item The Universe undergoes a thermal phase after reheating when the $\mathbb Z_2$ symmetry is approximately restored. 
\end{enumerate}

\section{Minimal model dynamics}
\label{sec: model}

In the previous section, we saw a first example of how the presence of two independent sources of $\mathbb Z_2$ symmetry breaking in the potential of the modulus $\phi$ can be used to realise a deterministic story for the evolution of $\phi$. 
In order to provide this symmetry breaking, however, the model made assumptions about the non-minimal coupling to gravity. 
In this section we will demonstrate that a deterministic model can already be successfully achieved with very minimal assumptions about the potential of $\phi$. As the most general potential of the modulus $\phi$ already contains two spurions of $\mathbb Z_2$ symmetry breaking, we generally do not need to invoke the non-minimal coupling.

We begin by considering the most general renormalisable potential for a modulus (real scalar field) with an explicitly broken $\mathbb Z_2$ symmetry:
\begin{equation}
    V(\phi) = \kappa_1^3\phi - \frac{\mu^2}{2}\phi^2+\frac{\kappa_3}{3!}\phi^3 + \frac{\lambda}{4!}\phi^4.
\label{eq:classical_potential}
\end{equation}
Here we fix the signs of $\lambda>0$ (such that the potential is bounded from below) and $\mu^2 > 0$, but allow the signs of $\kappa_1$ and $\kappa_3$ to vary. 
In this minimal scenario, the parameters $\kappa_1$ and $\kappa_3$ already provide the two independent sources of symmetry breaking needed, whereas the location of $\phi=0$ is again identified as the place where the new coloured fermions $q$ become massless. 
We make no additional assumptions about the origin of the symmetry breaking pattern, the size of the parameters or new couplings to other sectors, as long as $|\kappa_1|,|\kappa_3|\lesssim \mu$ so that there is an approximate $\mathbb Z_2$ symmetry. Instead, we explore all of the parameter space for this generic potential in which the criteria of our mechanism can be realised.

The potential (\ref{eq:classical_potential}) has the appearance of an asymmetric double well (see for example the blue curve in the top left panel of Fig.~\ref{fig: phi true}). 
One can determine exactly which of its minima is the true minimum by shifting the field by $\phi \to \phi - \kappa_3/\lambda$ to remove the cubic term, so that it takes the form
\begin{equation}
V(\phi) =   \left(\kappa_1^3+\frac{\kappa_3^3+3 \kappa_3 \lambda  \mu ^2}{3 \lambda ^2}\right)\phi-\frac{ \left(\kappa_3^2+2 \lambda  \mu ^2\right)}{2 \lambda }\frac{\phi^2}{2}+\frac{\lambda  \phi ^4}{4!}.
\label{eq:amaliashift un-rescaled}
\end{equation}
The sign of the linear coefficient is then opposite to the sign of the field value at the true minimum. 

At high temperatures following reheating, the classical potential $V(\phi$) for the scalar field is modified due to its interactions with the hot primordial plasma. 
These finite-temperature effects enable the $\mathbb  Z_2$ symmetry to be approximately restored at high temperatures. 
Like in the previous example, this approximate symmetry restoration erases all information in the potential about which minimum the scalar field lived in during inflation. 
The story of the subsequent evolution of the potential, however, differs from the previous toy model in one very important aspect. 
Previously, the potential contained only linear couplings in the field, meaning that the field would always settle in the true minimum after inflation. 
As we shall demonstrate, the presence of the cubic coupling $\kappa_3$ now makes it possible for the field to roll into either the true or false minimum. 
A large parameter space to end in either vacuum means that a deterministic model where the field lives in opposite minima during and after inflation can be easily achieved from this minimal potential.

\subsection{The evolution of the potential}
\label{sec: evolution pot}
In this subsection, we will study the form and evolution of the temperature dependent effective potential for the minimal model and identify the temperatures associated with transitional points in its behaviour.
We first simplify the analysis of the parameter space for $V(\phi)$ by rescaling the field $\phi$, the temperature $T$ and the four parameters of the potential in Eq.~\eqref{eq:classical_potential} into quantities which are dimensionless in units of both energy and couplings.
More details of this treatment are clearly summarised in \cite{Panico:2015jxa}. 
Whereas in natural units the Planck constant $\hbar$ is set to 1 and everything is expressed in units of energy, it can be convenient to restore $\hbar$ and work in units of lengths $L$ and couplings $C$ (of dimension $\hbar^{-1/2}$). 
In these units, masses and fields have different dimensionalities (respectively, $L^{-1}$ and $C^{-1}L^{-1}$), and the quartic coupling $\lambda$ has dimension $C^2$.%
\footnote{Temperatures have the same dimension as fields and the Planck mass, and the Boltzmann constant $k_B$ has dimension $C^{-1}$.}
We collect in Table~\ref{tab: rescalings} our definitions of the rescaled parameters (denoted by a tilde). 
The advantage of this procedure is that most of the physical behaviour in what follows, as shown in Figs.~\ref{fig: phi true}, \ref{fig: phi false} and \ref{fig: phi decayed}, will be encoded in $\kat$, $\kct$, $\yt$, and only occasionally on $\lambda$, whereas $\mu$ will not appear.
\begin{table}[h!] \centering
\begin{tabular}{cc}
\toprule
$\phit$ & $\phi\, \mu^{-1}\lambda^{1/2}$ \smallskip\\
$\Tt$ & $T\, \mu^{-1}\lambda^{1/2}$ \smallskip\\
$\yt$ & $y\, \lambda^{-1/2}$ \smallskip\\
$\kat^3$ & $\kappa_1^3\, \mu^{-3}\lambda^{1/2}$ \smallskip\\
$\kct$ & $\kappa_3\, \mu^{-1}\lambda^{-1/2}$ \\
\bottomrule
\end{tabular}
\caption{Notation for the model parameters, and definition of rescaled parameters which are dimensionless in units of energy and $\hbar$.}
\label{tab: rescalings}
\end{table}

The effect of finite temperature and quantum corrections on the classical potential \eqref{eq:classical_potential} are contained in the effective potential. 
The calculation of the effective potential at one-loop is discussed in detail in the Appendix \ref{app: V finite T}. 
We state here the final rescaled result
\begin{equation}
 V_{\text{eff}}(\phit, \Tt) = \frac{\mu^4}{\lambda}\Bigg[\left(\kat^3+\frac{1}{24} \kct \Tt^2\right)\phit +\left(-1 + \frac{\Tt^2}{24} \left(12 \yt^2+1\right)\right)\frac{\phit^2}{2} +\frac{\kct\phit^3}{3!} +\frac{\phit^4}{4!}\Bigg].
\label{eq:Vrescaled}
\end{equation}

At temperatures roughly higher than the mass of the scalar $\mu$, or $\Tt\gtrsim\lambda^{1/2}$ in terms of rescaled parameters, the effective potential has only one minimum.  
Once the Universe has cooled to a temperature we define to be $\Tcat$, the second minimum first appears. 
In the limit $\lvert \kat \rvert ,\lvert \kct \rvert \ll 1$, i.e.~soft $\mathbb Z_2$ symmetry breaking, this temperature approximately coincides with the temperature of spontaneous symmetry breaking, calculated when the mass term in the effective potential changes sign
\begin{equation}
\Tcat \simeq \frac{2 \sqrt{6}}{\sqrt{12 \yt^2+1}} \qquad \left(\lvert \kat \rvert ,\lvert \kct \rvert \ll 1\right) \,.
\end{equation}
With the inclusion of explicit symmetry breaking terms, the transformation of the potential from a single minimum to the asymmetric double well of the zero-temperature potential occurs in a qualitatively different manner from the conventional picture of spontaneous symmetry breaking. 
This behaviour of the potential, which can be seen on the left hand side of Figs.~\ref{fig: phi true} and \ref{fig: phi false}, ensures that by the time the second minimum appears, the scalar field has already rolled into the first minimum. 
Compared to the evolution of a spontaneously broken $\mathbb Z_2$ symmetry, we now not only have a low temperature potential where the two minima are no longer degenerate, but also a deterministic evolution into one of these asymmetric minima. 
A detailed analytical calculation of the temperature $\Tcat$ with significant explicit $\mathbb Z_2$ symmetry breaking is discussed in Appendix~\ref{app: 3TC1}.

We now identify a second important temperature in the evolution of the potential, relevant when the false vacuum appears first after reheating (see Fig.~\ref{fig: phi false}). 
We define the temperature $\Tcbt$ to occur when the energy densities of the two minima are degenerate. 
The subscript ${\rm c}$ here acknowledges that this is the critical temperature relevant for a first order phase transition from the false to true vacuum -- this possibility will be explained further in Sec.~\ref{sec:vacuumdecay}. 
To calculate $\Tcbt$, we first perform an identical shift to the effective potential to that of Eq.~(\ref{eq:amaliashift un-rescaled}), previously used to determine the true minimum of the classical potential. This is given in rescaled form by $\phit \to \phit -\kct$. 
Once again this removes the cubic term:
\begin{equation}
   V_{\text{eff}}(\phit,\Tt)= \frac{\mu ^4}{\lambda} \left[ \left( \kat^3+\kct \left(1+\frac 13 \kct^2-\frac 12 \Tt^2 \yt^2\right)\right)\phit+\left(-1-\frac{\kct^2}{2}+ \frac{\Tt^2}{24} \left(12 \yt^2+1\right)\right)\frac{\phit^2}{2}+\frac{\phit^4}{4!}\right].
\label{eq:shiftedlinearcoeff}
\end{equation}
We note that there is a particular value of $\Tt$ for which the linear term vanishes in the shifted potential. 
At this temperature, the potential is temporarily $\mathbb Z_2$ symmetric and the two minima must lie at the same depth.
We find this to be
\begin{equation}
\Tcbt = \frac{\sqrt{2}}{\yt}\sqrt{\frac{\kct^2}{3}+\frac{ \kat^3 }{\kct}+1}\,.
\label{eq: Tc}
\end{equation}
In the limit of a vanishing Yukawa coupling, we see that $\Tcbt$ diverges, indicating the necessity of Yukawa couplings in enabling the scalar field to roll into the false vacuum after inflation.

\subsection{Vacuum selection}
\label{sec:vac selection}
\begin{figure}[t] \centering
{Field $\phi$ settling in \textbf{false} vacuum (point \includegraphics[height=1em]{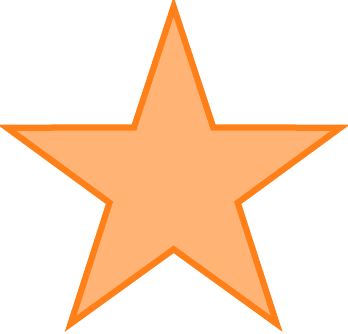} in Fig.~\ref{fig:MoneyPlot})} \smallskip\\
\includegraphics[width=.8\textwidth]{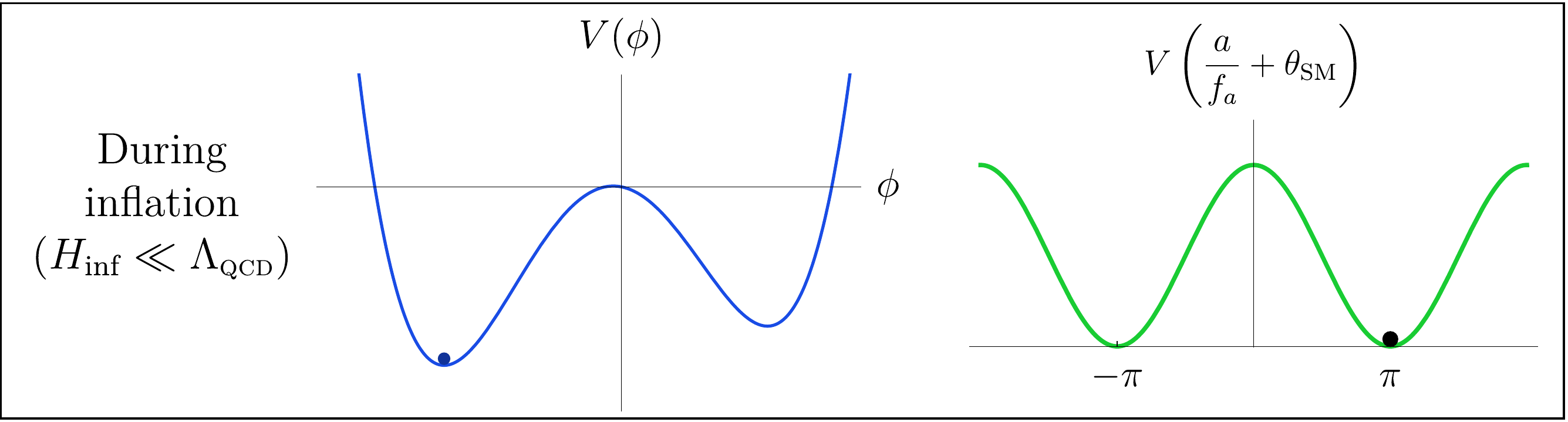}\smallskip\\%
\includegraphics[width=.8\textwidth]{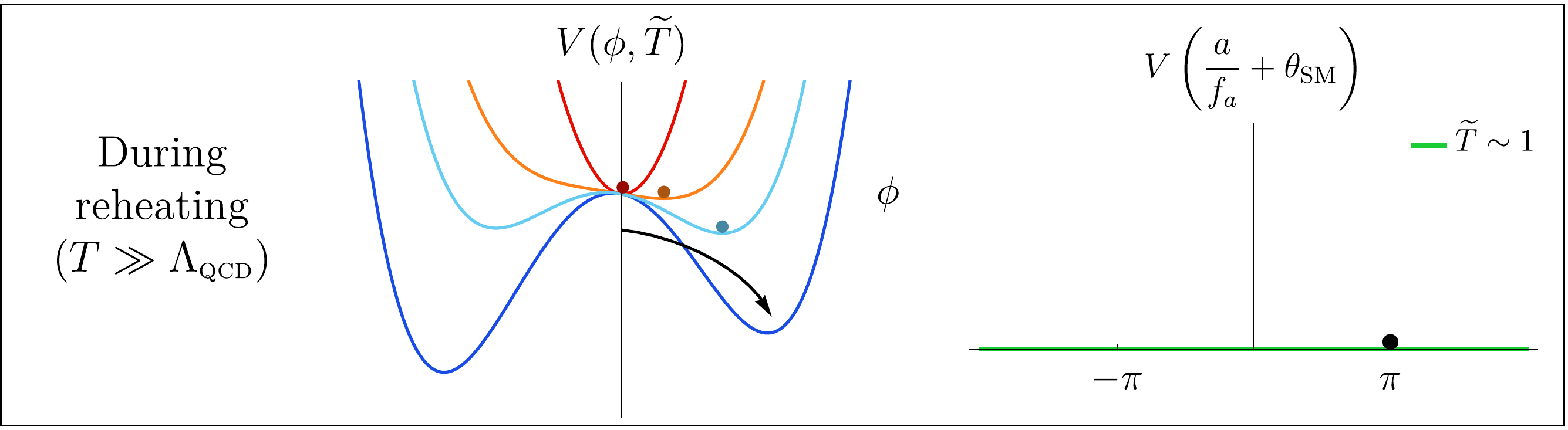} \smallskip\\%
\includegraphics[width=.8\textwidth]{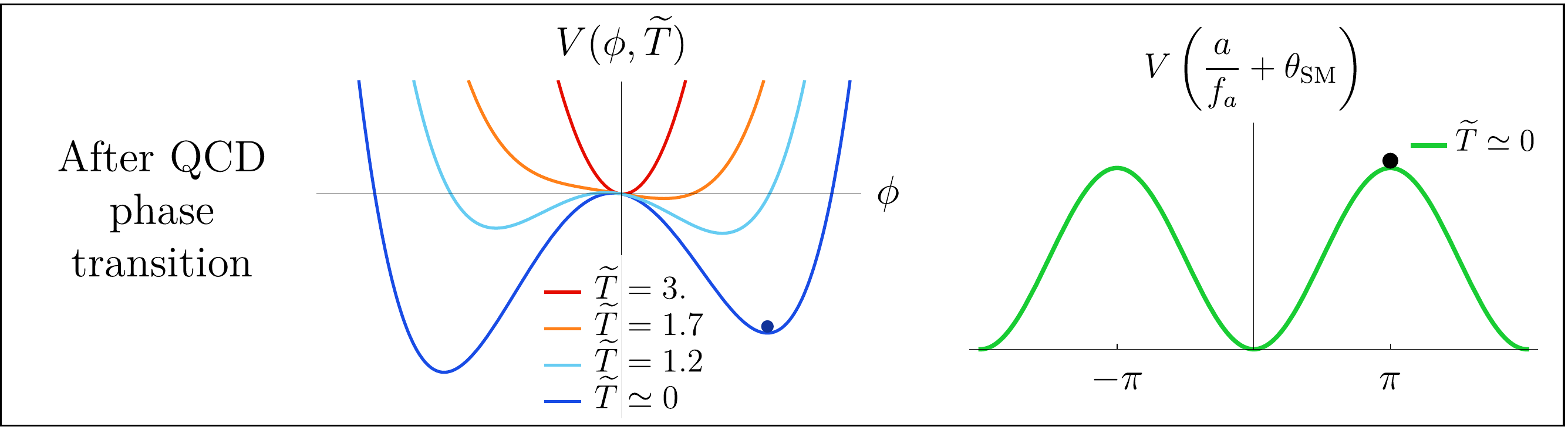} \smallskip\\%
\caption{Figure analogous to Fig.~\ref{fig: phi true}, for the case in which $\phi$ lies in its true vacuum during inflation and ends up in the false vacuum today.
For visualisation purposes, the Yukawa coupling $y$ is taken to be positive in these figures. Notice that in this case, when $y>0$, ${\rm arg} [m_q] = \pi$ during inflation.
}
\label{fig: phi false}
\end{figure}

\begin{figure}[h!] \centering
{Field $\phi$ settling in true vacuum after thermal decay (point \includegraphics[height=1em]{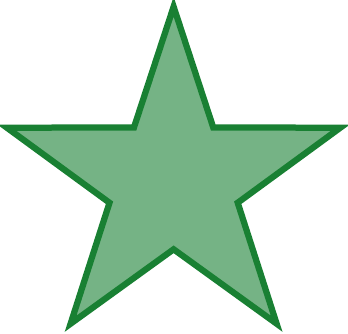} in Fig.~\ref{fig:MoneyPlot})} \smallskip\\
\includegraphics[width=.8\textwidth]{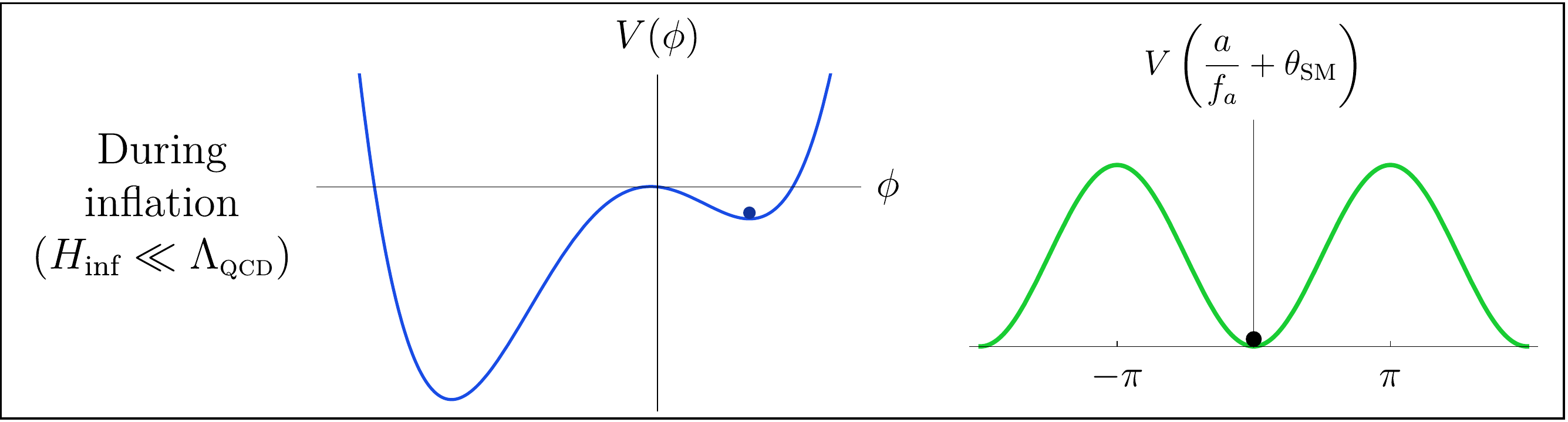}\smallskip\\%
\includegraphics[width=.8\textwidth]{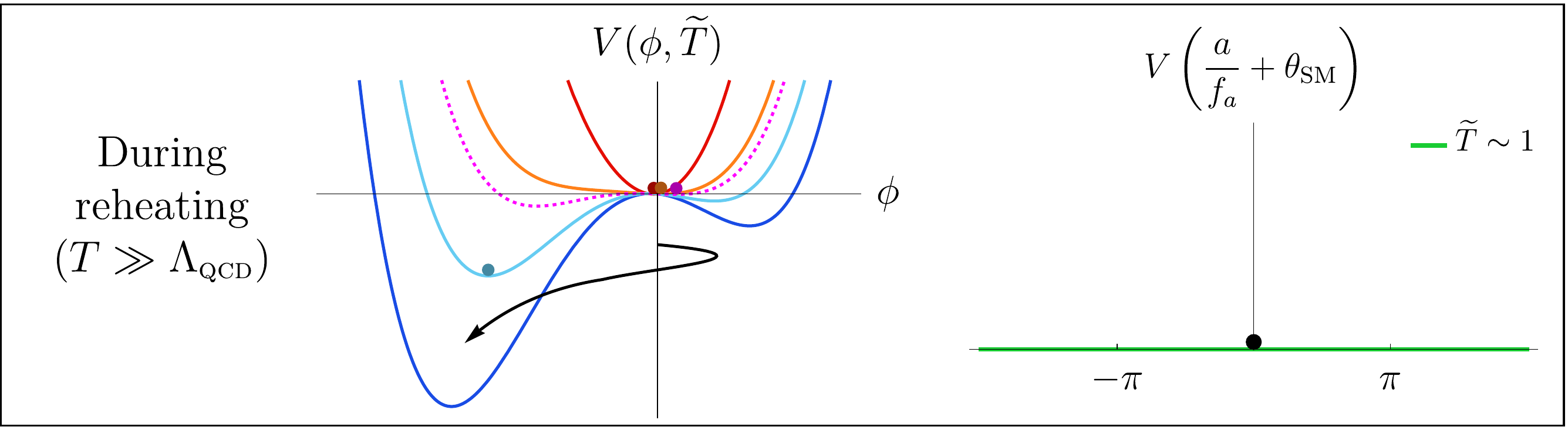} \smallskip\\%
\includegraphics[width=.8\textwidth]{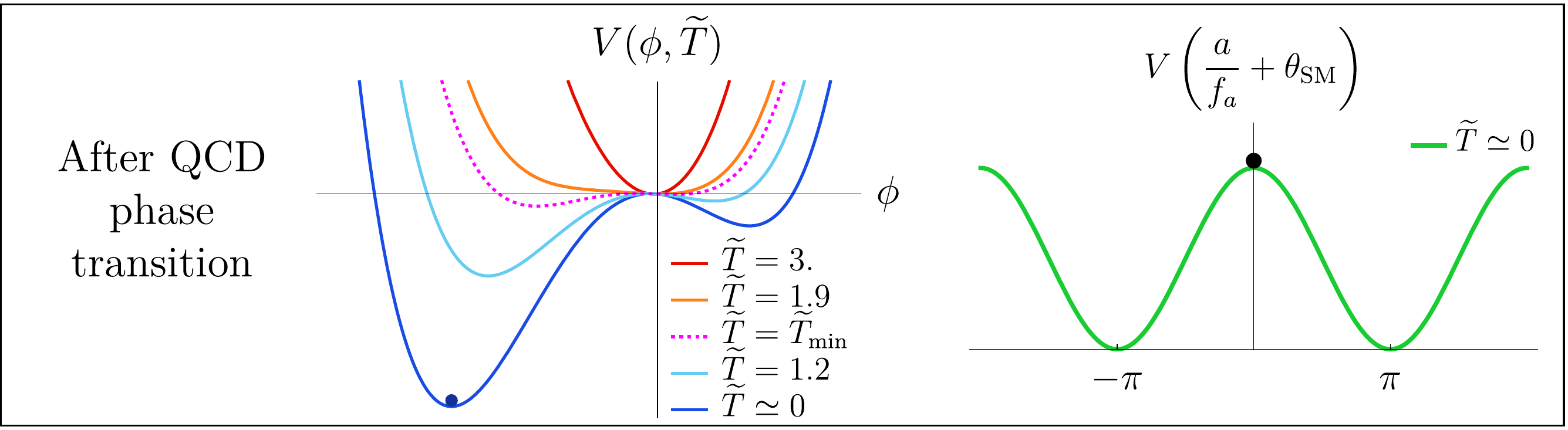} \smallskip\\%
\caption{Figure analogous to Figs.~\ref{fig: phi true} and \ref{fig: phi false}, for the case in which $\phi$ lies the false vacuum during inflation and rolls to the false vacuum at the beginning of the thermal phase. 
At a temperature around $\Tt_\text{min}$ (see Fig.~\ref{fig: thermal bounce}), for which $V(\phi,\Tt_\text{min})$ is marked with a dashed magenta line, the false vacuum is unstable against thermal decay, and $\phi$ ends up in the true vacuum today.
Also in this case, the VEV of $\phi$ changes sign between the inflationary time and today.}
\label{fig: phi decayed}
\end{figure}

We will now discuss a procedure to determine exactly which minimum, true or false, the field rolls into after inflation. 
This period of the evolution of $\phi$ is visible in the central plots in Figs.~\ref{fig: phi true}, \ref{fig: phi false} and \ref{fig: phi decayed}.
To begin, we point out that while one might expect $\mathbb Z_2$ symmetry to be fully restored in the limit $\Tt\to\infty$, this is actually disguised in the potential in Eq.~\eqref{eq:Vrescaled} by the presence of a thermal tadpole contribution, which shifts the position of the minimum at high temperatures away from $\phit=0$. We therefore perform a shift in the field that returns the high-temperature minimum asymptotically to the origin
\begin{equation}
\phit \to \phit-\frac{\kct}{12\yt^2+1}.
\end{equation}
After performing the above shift, the derivative of the potential with respect to $\phit$ evaluated at zero is a temperature-independent constant
\begin{equation}
\frac{\mathrm d V}{\mathrm d \phit} \bigg|_{\phit=0} =\frac{\mu^4}{\lambda}\left[\kat^3+\frac{\kct^3(18\yt^2+1)}{3(12\yt^2+1)^3}+\frac{\kct}{12\yt^2+1}\right]\,.
\label{deriv0}
\end{equation}
The sign of this quantity determines which side of the origin the field will roll towards after inflation. We can justify this as follows: at very high temperatures, the field will be located asymptotically  close to $\phit=0$ in the shifted potential. 
Now, as the slope at the origin is temperature independent, the field will roll to the side determined by the sign of Eq.~\eqref{deriv0}, and will be prevented at all temperatures from classically rolling to the other side of the origin. 
To determine whether the side where the field is restricted corresponds to the true or false vacuum at low temperatures, we simply compare the sign of Eq.~\eqref{deriv0} to the sign of field value of the true or false vacuum of the zero-temperature potential (see Eq.~\eqref{eq:amaliashift un-rescaled}). 
For example, if the sign of Eq.~\eqref{deriv0} is positive and the zero-temperature true minimum is located at a negative value of $\phit$, then we conclude that the field must end in the true vacuum.  

The rolling behaviour of the field is particularly simple in the soft symmetry breaking limit ($\lvert \kat\rvert\sim \lvert \kct\rvert\ll 1$), where it is fully determined by the cubic coupling $\kct$. 
In this simple case the field always rolls into the true vacuum after inflation, because the absolute minimum of the potential is determined only by $\kct$. 
The general case for arbitrary $\lvert\kat\rvert,\lvert\kct\rvert \leq 1$ is presented in Fig.~\ref{fig:MoneyPlot}, up to the detail of false vacuum decay, which will be discussed later in Sec.~\ref{sec:vacuumdecay}. 
The lines that enclose the regions of a false vacuum final state are the solutions of $\left. \tfrac{\mathrm d V}{\mathrm d \phit} \right|_{\phit=0}=0$.

\subsection{Vacuum decay}
\label{sec:vacuumdecay}
If the field rolls into the false vacuum after inflation, it is susceptible to decay to a true vacuum final state, which can reduce the parameter space available to end in the false vacuum. In this subsection we will see that while zero-temperature quantum tunnelling processes do not affect any of the parameter space of our model, an analysis of finite-temperature effects shows that a small but significant part of the parameter space actually decays due to thermal fluctuations. 
For the sake of simplicity we estimate the vacuum decay rate using the thin-wall approximation (TWA) \cite{Coleman:1977py}. 
This approximation, when applicable, offers a good analytical control of the decay rate as a function of few parameters.  

\subsubsection{Zero-temperature tunnelling}\label{zero_T_decay}
At zero temperature, false vacuum decay is a tunnelling process that occurs via the nucleation of bubbles of true vacuum by quantum fluctuations within the false vacuum. A first order phase transition can then occur through the expansion of these bubbles. We estimate the probability $\Gamma$ of formation of a bubble of true vacuum in the false vacuum per unit spacetime volume $V$ as
\begin{equation}
\Gamma/V=Ae^{-B}\,.
\label{eq:0Tdecayrate}
\end{equation}
Here $B\equiv S_4$ is the Euclidean action for the bounce solution (the Euclidean bubble), which is $O(4)$ symmetric. The prefactor $A$ can be estimated as \cite{Linde:1981zj}:
\begin{equation}
A\sim \left(\frac{S_4}{2\pi}\right)^2 \mu^4\,.
\end{equation}
In the limit that the energy difference between vacua (bias) is a small parameter, the Euclidean action $B$ may be represented in closed form using the thin-wall approximation (TWA), as first described in \cite{Coleman:1977py}. 
We include the details of the TWA calculation for our setup in Appendix~\ref{app: vacuum decay}. The result for $B$ is found as:
\begin{equation}
B=\frac{27 \pi^2 S_1^4}{2 \epsilon^3}=
 \frac 1\lambda \frac{81\sqrt{3}\pi^2(\kct^3+2)^{9/2}}{(3\kat^3+\kct^3+3\kct)^3}\,
\end{equation}
where $S_1$ is the one-dimensional action and $\epsilon$ is the bias.
The probability of false vacuum decay at zero temperature is obtained by multiplying Eq.~\eqref{eq:0Tdecayrate} by the present spacetime volume $V\sim H_0^{-4}$. 
The false vacuum does not decay at zero temperature, that is 
\begin{equation}
H_0^{-4}A \, e^{-B}\ll 1\,,
\end{equation}
as long as $\lambda \lesssim 5$ and, therefore, we conclude that the false vacuum region is stable against quantum decay processes in all of its parameter space.

\subsubsection{Finite-temperature vacuum decay}
\label{subsec:Thermalvacuumdecay}

At high temperatures, false vacuum decay no longer occurs via quantum tunnelling but by thermal jumps over the barrier, where bubbles of true vacuum are nucleated by thermal fluctuations. 
This can occur at temperatures below $\Tcb$, defined previously in Eq.~\eqref{eq: Tc} as the moment when the energy density of the true vacuum falls enough such that it becomes degenerate with the false. 
The associated rate is suppressed by the Maxwell-Boltzmann factor, $e^{-\beta E_b}$:
\begin{equation}\label{thermal_rate}
\Gamma(T)/V(T)=A(T)e^{-\beta E_b(T)}\,,
\end{equation}
where $\beta=1/T$ and $E_b$ is the free-energy, or equivalently the three-dimensional action $S_3$ of the unstable bubble. 
This formula is valid as long as $E_b \gg T$, i.e.~when the probability of bubble nucleation is small. 
Following the procedure in Appendix~\ref{app: vacuum decay} where we again apply the TWA, we find:
\begin{equation}\label{free_energy}
\frac{E_b(T)}{T}=\frac{128\pi}{243\sqrt{3}}\frac{ v(T)^9\lambda^{3/2}}{T\epsilon(T)^2}\,,
\end{equation}
with the temperature dependent VEV and bias given by
\begin{equation}\label{eq:thermal_bias_and_VEV}
\begin{gathered}
v(T)^2=\frac{\mu^2}{\lambda}\left[6+3\kct^2-\frac{\Tt^2(1+12\yt^2)}{4}\right]\,,\\
\epsilon(T)= \frac{\mu^3}{\sqrt{\lambda}}\yt^2\kct\left[\Tcbt^2-\Tt^2\right]v(T)\,.
\end{gathered}
\end{equation}
For the thermal case, the prefactor can be approximated by $A(T)\sim T^4(\frac{E_b}{2\pi T})^{3/2}$ \cite{Linde:1981zj}. 

Assuming that at the time of phase transition the Universe is radiation dominated and the temperature is around the TeV scale,
\begin{equation}
V(T)A(T)e^{-E_b(T)/T}\approx 1 \quad \text{for}\quad \frac{E_b(T)}{T}\sim 140\,.
\end{equation}
At the moment of formation, the radius of a true vacuum bubble is much smaller than the Hubble length, $H^{-1}$. 
This is related to the smallness of the nucleation temperature (around the TeV scale for our case) compared to the Planck scale. 
As a result, the phase transition occurs due to the nucleation and subsequent expansion of several bubbles. 
The bubble walls reach semi-relativistic speeds in a short time and a handful of bubbles is sufficient to convert all of space into a true vacuum state in a few Hubble times \cite{Anderson:1991zb}.
\begin{figure}[h!] \centering
\includegraphics[width=.6\textwidth]{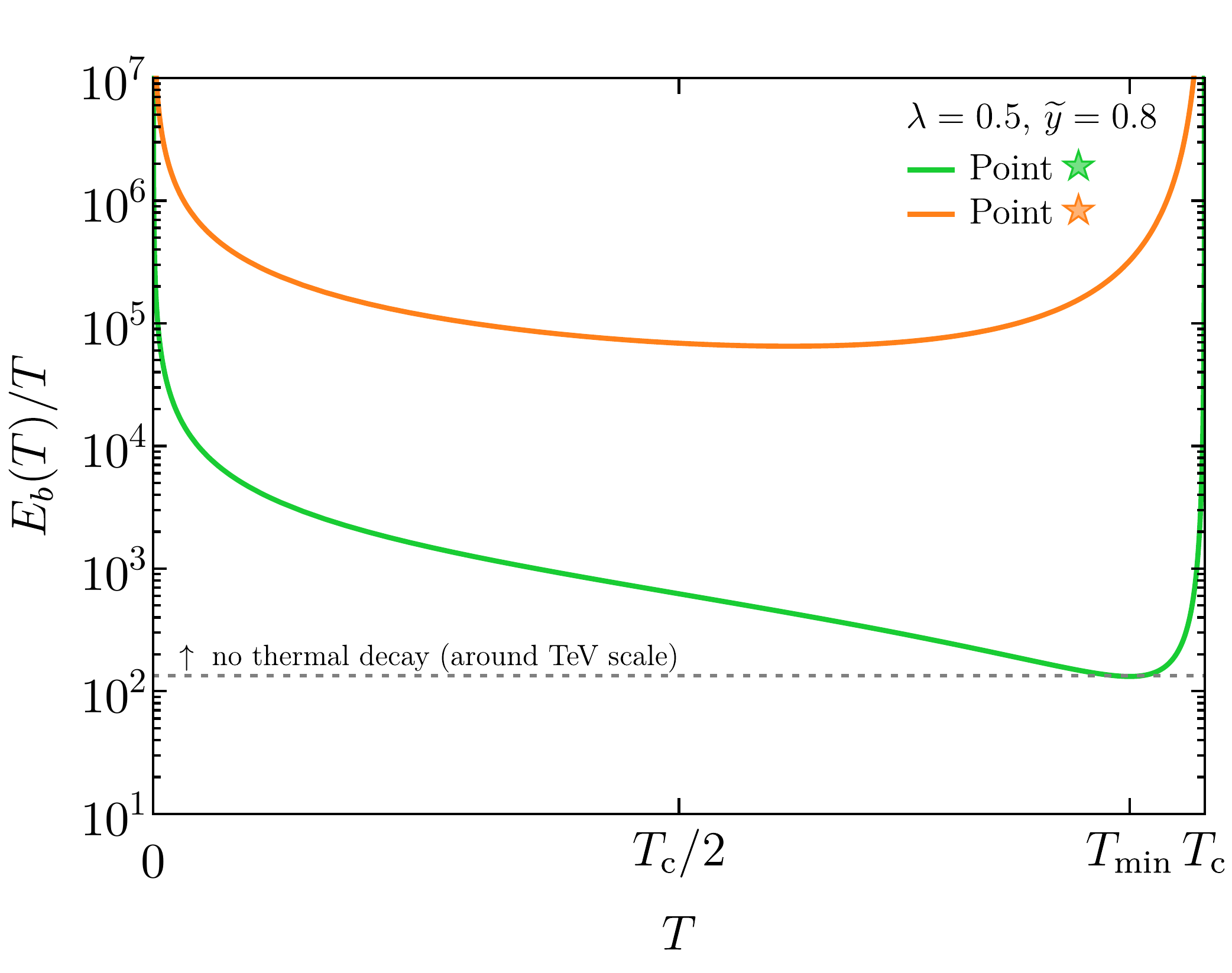}
\caption{Temperature dependence of the exponent for the probability of thermal decay, given by the ratio of free energy of the bubble $E_b(T)$ and temperature as in Eq.~\eqref{free_energy}.
The two curves correspond to the two points of the parameter space marked with orange and green stars in Fig.~\ref{fig:MoneyPlot}.
The orange line corresponds to a case where the false vacuum is stable (in our case, $E_b(T)/T\gg 140$) for all temperatures. 
For the green line, instead, the minimum of $E_b(T)/T$ at $T_\text{min}$ is below the threshold and the green point of Fig.~\ref{fig:MoneyPlot} can be considered unstable against thermal decay throughout the thermal history of the Universe. 
It can be seen for both cases that, as $T$ approaches $\Tcb$ (defined in Eq.~\ref{eq: Tc}), the energy difference between the two vacua vanishes and the decay rate goes to zero.}
\label{fig: thermal bounce}
\end{figure}

Following the procedure detailed in Appendix~\ref{app: vacuum decay} we determine the stability of the false vacuum region by imposing that the minimum of the free energy in Eq.~\eqref{free_energy} is larger than the threshold required to nucleate the bubbles
\begin{equation}
\frac{E_b(T_\text{min})}{T_\text{min}}> 140\,.
\end{equation}
Here $T_\text{min}$ is the temperature at which the semi-classical exponent is minimised. In Fig.~\ref{fig: thermal bounce} we evaluate $E_b(T)/T$ for two points in the plane $(\kat,\kct)$ with qualitatively different behaviour: stable and unstable false vacuum.

We note that only a small part of the false vacuum region decays into true vacuum due to thermal fluctuations, see green region in Fig.~\ref{fig:MoneyPlot}. 
For completeness we also include the validity condition for the TWA. 
This approximation is valid when the radius of the critical bubble, $r_c$, is larger than the width of the wall: $\Delta r\sim (V_\text{min}'')^{-1/2}$, where $(V_\text{min}'')^{-1/2}$ stands for the curvature of the potential at the minimum and primes $^\prime$ denote derivatives with respect to $\phi$ (see Appendix~\ref{app: vacuum decay} for details). 
We have chosen, as a threshold, $r_c/\Delta r >4$. 
Note that all the stable false vacuum region is well within the region where the TWA is valid, supporting the consistency of our results.

\subsection{Results}
\begin{figure}[h!] \centering
\includegraphics[width=.6\textwidth]{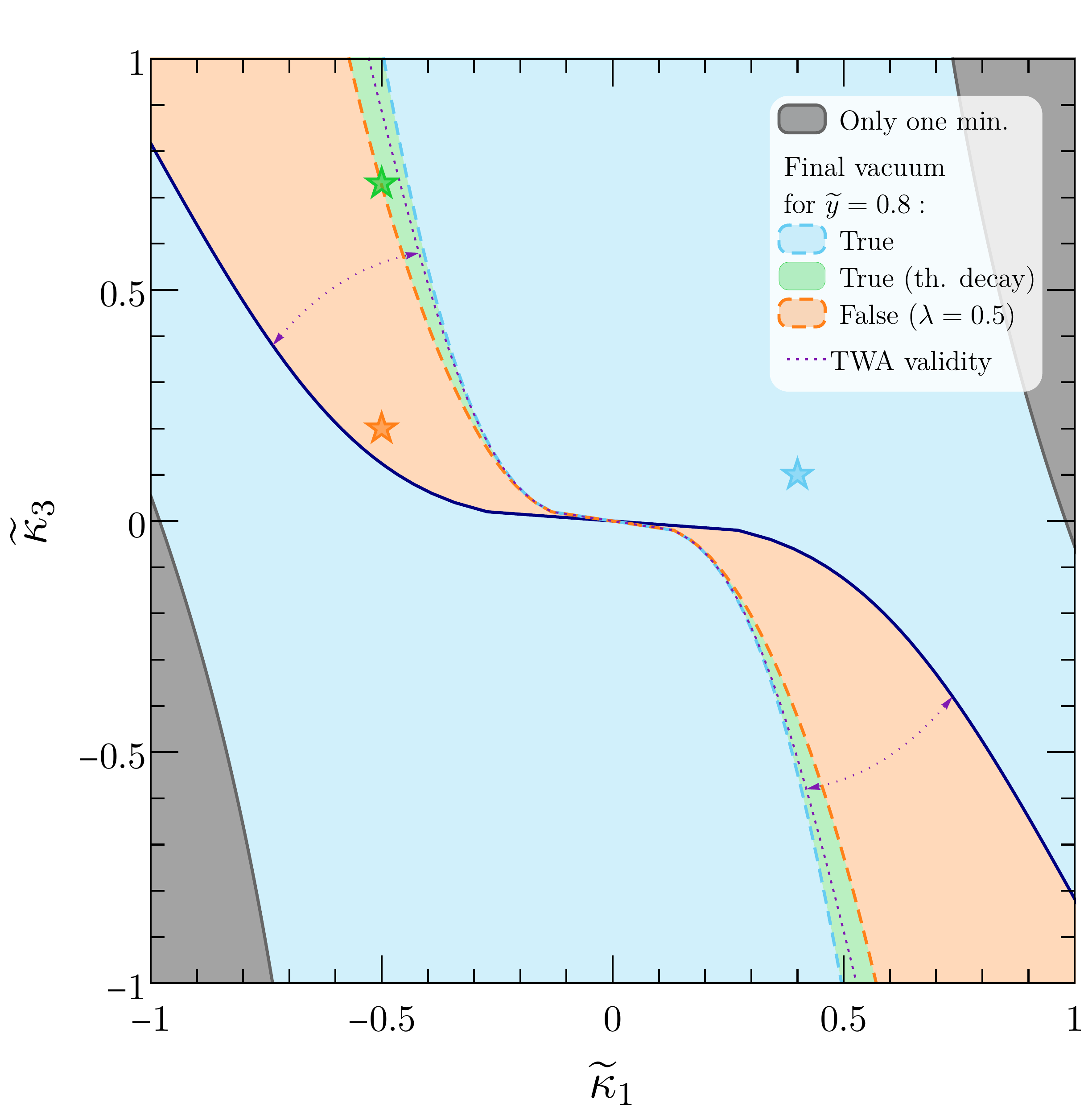}
\caption{Parameter space for the field $\phi$ where it can settle either to its true vacuum (blue and green regions) or false vacuum (orange region) in the thermal history of the Universe.
The parameters $\kat$, $\kct$, $\yt$, $\lambda$ are defined in Eq.~\eqref{eq:classical_potential} and Table~\ref{tab: rescalings}.
In the regions which are not shaded in grey in this plot, $V(\phi)$ displays two non-degenerate minima.
Depending on the solution to the measure problem, $\phi$ during inflation was either in its false or in its true vacuum.
If $\phi$ resided in its false vacuum during inflation, then the blue and green regions of this plot realise the mechanism shown respectively in Fig.~\ref{fig: phi true} and \ref{fig: phi decayed} (for the points marked respectively by blue and green stars) and $\phi$ eventually settles in the true vacuum, which has an opposite sign with respect to inflationary one.
If instead $\phi$ was in its true vacuum during inflation, then the orange regions of this parameter space realise the evolution shown in Fig.~\ref{fig: phi false} (corresponding to the point marked by the orange star).
Therefore, independently of the solution to the measure problem there is a fraction of order one (orange region, 20\% in this figure) of the parameter space of $\phi$ where the field $\phi$ can start from the lower energy minimum during inflation and ends in the false vacuum today, and another fraction of order one (blue and green region, 80\% in this figure) of the parameter space of $\phi$ where the field $\phi$ can start from the higher energy minimum during inflation and ends in the true vacuum today.
In all cases, that is, all coloured regions, the axion field has an initial misalignment angle close to $\pi$.
The difference between blue and green regions is that, in the latter case, $\phi$ was initially rolling to the false vacuum during the thermal phase, but then thermally decayed to the true vacuum, as shown in Fig.~\ref{fig: phi decayed}. 
This calculation is performed using the thin-wall approximation (TWA), whose regime of validity is shown by the purple dotted arrows in this plot.
}
\label{fig:MoneyPlot}
\end{figure}
We are now ready to present in Fig.~\ref{fig:MoneyPlot} our final results for the regions in the parameter space of $\phi$ that realise the evolution described in the previous sections.
In this section and in the figures we take the Yukawa coupling $y$ to be positive for illustrative purposes.

As discussed in Sec.~\ref{sec: evolution pot}, it is useful to rescale scales and couplings in units of energies and $\hbar$ so that we can isolate the dependence of most of the relevant physical quantities on the dimensionless parameters $\kat$, $\kct$, $\yt$ and occasionally $\lambda$ (see Table~\ref{tab: rescalings} and Eq.~\eqref{eq:Vrescaled}).
We restrict ourselves to the range $[-1,1]$ for the parameters $\kat$ and $\kct$, motivated by the approximate $\mathbb Z_2$ symmetry.
As for the rescaled Yukawa coupling $\yt$ and the quartic $\lambda$, we focus on values smaller than about $1$, to ensure the validity of the high-temperature expansion for the study of the thermal evolution and the thin-wall approximation for the vacuum decay rate. 
Larger values of $\yt$ and hence $\lambda$ can also be considered within a more extended numerical study, which lies beyond the scope of this work.

We hatch in dark grey the regions where zero temperature potential $V(\phi,\Tt=0)$ displays only one minimum and does not satisfy the necessary condition of the existence of two minima.
For parameters in the light blue regions, the field $\phi$ ends in its true vacuum at low temperatures. This region corresponds to the evolution of $\phi$ shown in Fig.~\ref{fig: phi true}. 
In the orange and green regions, $\phi$ starts rolling towards its false vacuum after the end of inflation, corresponding respectively to Figs.~\ref{fig: phi false} and \ref{fig: phi decayed}.
These regions are found by using Eq.~\eqref{deriv0}.
In the green region, however, the false vacuum of $V(\phi,\Tt)$ is unstable against vacuum decay (see Fig.~\ref{fig: thermal bounce}), and at the temperature $\widetilde T_\text{min}$ shown in dashed magenta in Fig.~\ref{fig: phi decayed}, $\phi$ thermally decays to the true vacuum.

The key information of Fig.~\ref{fig:MoneyPlot} is the fraction of parameter space covered in orange (where $\phi$ starts in the lower energy minimum during inflation and ends in the false vacuum) relative to the blue and green regions (where $\phi$ starts in the higher energy minimum during inflation and settles in the true vacuum at later times). In all cases, the axion field has an initial misalignment angle close to $\pi$.
Depending on the measure, the field $\phi$ preferentially lives in either the higher energy minimum or the lower energy minimum during inflation. 
The possibility of flipping the sign of the VEV of $\phi$ after the end of inflation is the core of this mechanism that circumvents the measure problem.
The probability of ending in the false vacuum relative to the true vacuum is the percentage number quoted in the caption of Fig.~\ref{fig:MoneyPlot}, and it quantifies the tuning in our scenario.
\begin{figure}[h!]\centering
\includegraphics[width=.49\textwidth]{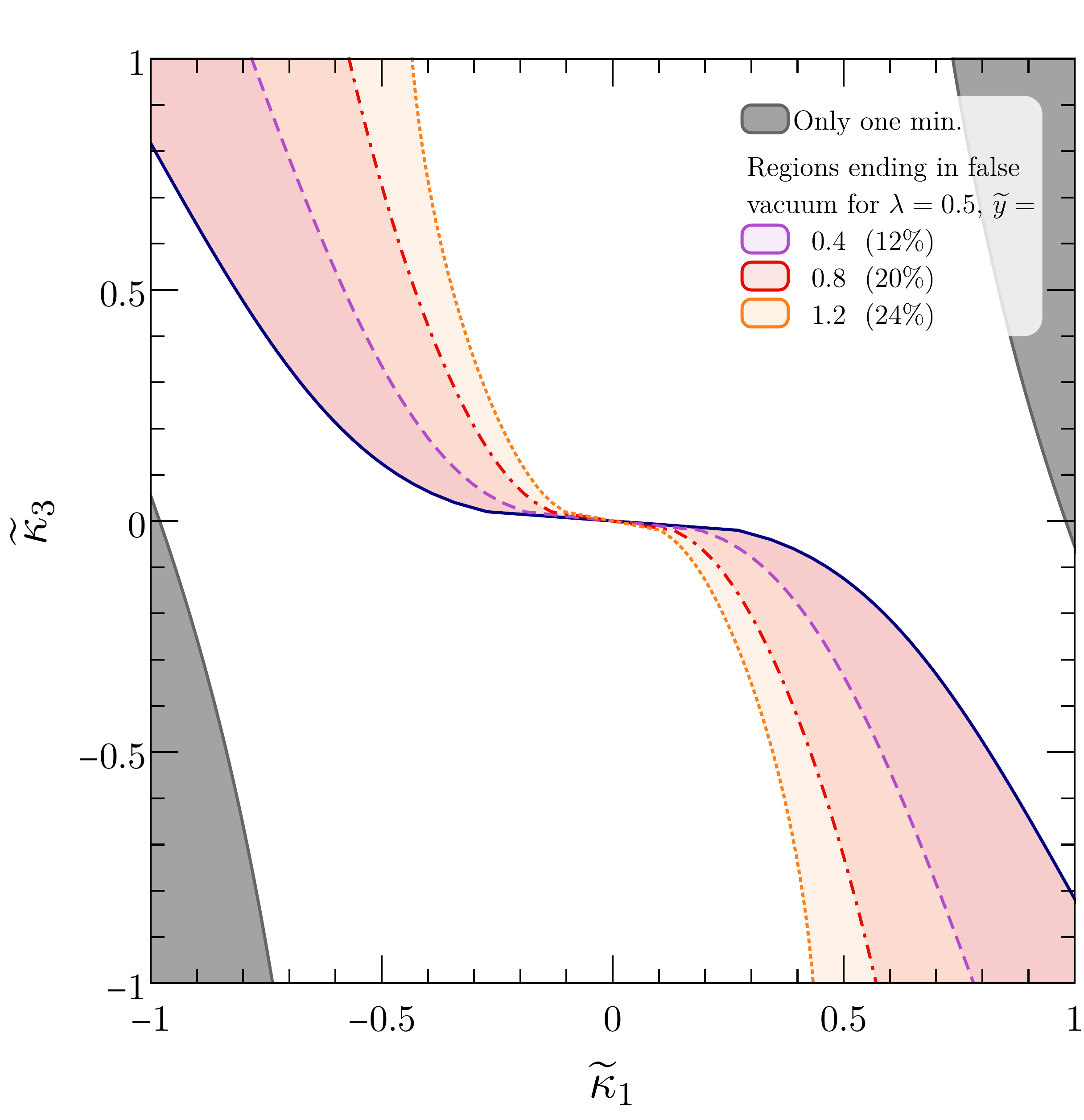} \hfill
\includegraphics[width=.49\textwidth]{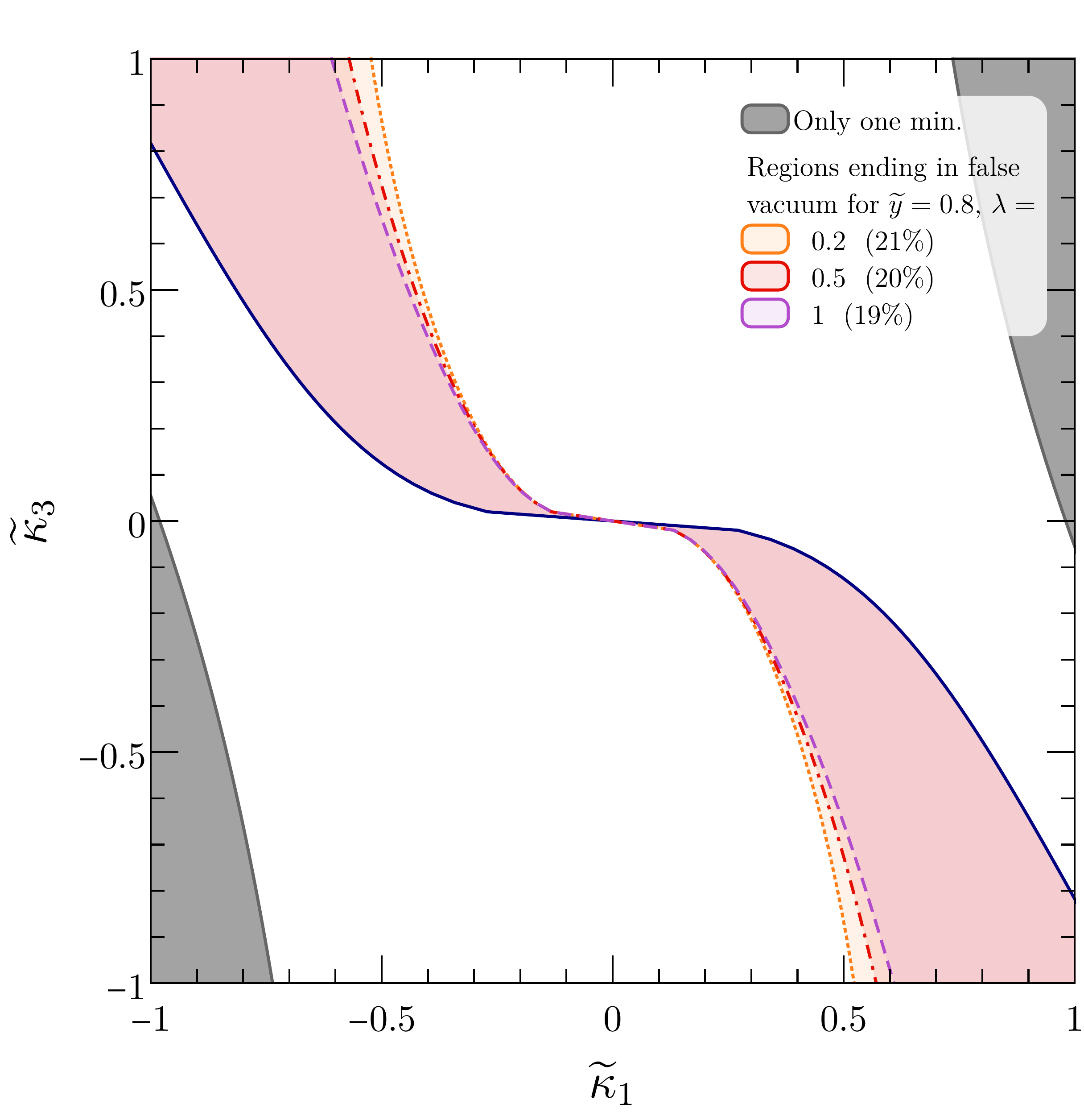}
\caption{Similarly to Fig.~\ref{fig:MoneyPlot}, we show the  regions where $\phi$ ends up in its false vacuum after the thermal phase (i.e.~the equivalent of the orange regions in Fig.~\ref{fig:MoneyPlot}) for different values of $\yt\sim 0.1-1$ and $\lambda\sim 0.1-1$.
Across the ranges shown in these plots, the shaded regions cover about 10\% to 20\%  of parameter space. 
This implies a modest amount of tuning in parameter space needed to realise the mechanism shown in Figs.~\ref{fig: phi true} and \ref{fig: phi false}, so that the axion field has an initial misalignment close to $\pi$.
}
\label{fig:MoneyPlot Yukawas}
\end{figure}

The mechanism works for a wide range of parameters $\yt$ and $\lambda$, as shown by Fig.~\ref{fig:MoneyPlot Yukawas} where we quote the tuning for different values.
In these rescaled units, we can isolate the dependence on the relevant parameters, and highlight that in the $(\kat,\kct)$ plane the probability of rolling to either the false or true vacuum depends on $\yt$, whereas $\lambda$ determines the probability of thermal decay (green region in Fig.~\ref{fig:MoneyPlot}).
For the values of  $\yt$ and $\lambda$ that we consider, the high-temperature expansion is a good approximation (see App.~\ref{app: V finite T}).

In summary, we present a mechanism that circumvents the measure problem with mild additional tuning (of order 20\%). 
In principle, our model needs a UV completion to solve the hierarchy problem associated to the real scalar field $\phi$, analogously to that of the SM Higgs.
For example, it is easy to embed this model in an extra-dimensional set-up with an $\mathcal O(100 \text{ TeV})$ fundamental scale, where the presence of an abundance of moduli fields is a natural consequence \cite{ArkaniHamed:1998rs}.

\subsection{Further checks on the parameter space}
The parameters of the scalar potential in Eq.~\eqref{eq:classical_potential} are subject to different requirements. 
Before moving to the phenomenological implications of the mechanism, in this subsection, we comment on the allowed parameter space from a purely theoretical point of view. 

\begin{enumerate}[(i)]
\item One of the fundamental prerequisites of the mechanism is reheating the $\phi$ sector. This means that after inflation, the interaction with the hot primordial plasma has to be efficient in order to approximately restore the $\mathbb Z_2$ symmetry by thermal effects. There are two main channels to thermalise the scalar field:
\begin{equation}
\phi\phi\leftrightarrow\phi\phi\,,\qquad 
\phi\phi\leftrightarrow \overline q q\,,
\end{equation}  
where $q$ stands for the new coloured fermion. Demanding that the rate of the above reactions is faster than cosmic expansion sets lower bounds to the quartic scalar self-coupling and the Yukawa coupling. 
A straightforward estimation of the rates gives
\begin{equation}
\lambda\gg (\TRH/\MP)^{1/2}\,,\qquad y\gg(\TRH/\MP)^{1/4}\,.
\end{equation}
These requirements are very easy to satisfy for reheating temperatures $\TRH \sim\mathcal{O}(1-10 \text{ TeV})$. Strictly speaking, we only need thermalisation to occur before the vacuum selection described in Sec.~\ref{sec:vac selection} around a temperature comparable to $\Tca$, which implies an even weaker requirement.
\item The second requirement is related to the size of the linear term in the shifted base, i.e., the derivative evaluated at $\phi\rightarrow 0$, see Eq.~\eqref{deriv0}. Deterministic vacuum selection after inflation demands that the field classically rolls to the first minimum relatively fast
\begin{equation}
V'(0)\gg v(T)\HI^2\,.
\end{equation}
The above inequality is easily satisfied in almost all the parameter space due to the small value of $\HI$  or, in other words, due to the smallness of $\TRH$ compared to the Planck mass, see Eqs.~\eqref{initial_misalignment} and \eqref{RH_temp}. 
As one would expect if the derivative vanishes, this is exactly on the false vacuum lines in Fig.~\ref{fig:MoneyPlot} (or arbitrarily close to them, where $V'(0)\rightarrow 0$) and in the exact $\mathbb Z_2$ limit, $\kappa_1=\kappa_3=0$, the condition above is not satisfied and we recover the classic situation where the choice of vacuum is stochastic. 
\item It is well known that the expansion of false vacuum bubbles is thermodynamically forbidden. However, they might alter our previous conclusions under certain circumstances. 
In particular when the final state is a false vacuum, any false vacuum bubble that materialises between $\Tca$ and $\Tcb$ becomes a true vacuum bubble after $\Tcb$ and, therefore, might expand indefinitely. 
Since our bubble calculation of thermal vacuum decay only applies for temperatures below $\Tcb$ (see Sec.~\ref{subsec:Thermalvacuumdecay}) we need to impose that in the case that one of these false vacuum bubbles appears between $\Tca$ and $\Tcb$ its lifetime is much shorter than the Hubble time associated to $\Tcb$. 
In this case, such a false vacuum bubble would collapse before becoming a true vacuum bubble, restoring the validity of our previous analysis. 
To this end, we impose that the characteristic lifetime of any false vacuum bubble, estimated as $\tau_\text{bubble}\sim\frac{\sigma_\text{wall}}{\Delta V}$, is much shorter than $H^{-1}(\Tcb)$. 
The surface tension of the bubble can be estimated as $\sigma_\text{wall}\sim\sqrt{\lambda}v(T)^3$ while the bias reads $\Delta V=\epsilon(T)$. Taking the expressions in Eq.~\eqref{eq:thermal_bias_and_VEV}, we get a very weak lower bound for the cubic term
\begin{equation}
\kappa_3\gg \frac{\lambda^{3/2}}{y^2}\frac{v^2(T)}{\MP}\,,
\end{equation}   
for any temperature $T$ between $\Tca$ and $\Tcb$. 
\end{enumerate}

\section{Phenomenology}
\label{sec: pheno}

In this section we briefly comment on the phenomenological implications of the mechanism and different bounds to the new coloured fermion and scalar fields. These implications can be classified as model independent and model dependent. The former arise solely from the large initial misalignment condition $\theta_i\approx \pi$ and have been extensively studied in \cite{Arvanitaki:2019rax}. The latter are related to the particular UV completion of the mechanism and, in our case, are related to the vector-like quark and the real scalar field.

\subsection{Model independent predictions}
As anticipated in the introduction, the large-misalignment mechanism produces a delay in the onset of axion oscillation around the minimum. This scenario has the following main consequences.

\begin{enumerate}[(i)]
\item \textbf{Small and dense femto-halos with a large number density}. 
These large-misalignment axion structures are much more abundant than standard axion miniclusters and can therefore significantly affect haloscope searches (see \cite{Arvanitaki:2019rax} for details). 
The underlying formation mechanism is quite different. 
Axion miniclusters originate due to the collapse of large primordial fluctuations of the axion field, requiring post-inflationary PQ symmetry breaking. 
Axion femto-halos arise when the PQ symmetry is not restored during reheating and the initial misalignment angle is close to $\pi$. In this case, density perturbations are enhanced due to parametric resonance effects.

\item \textbf{Isocurvature fluctuations.}
Another potential signature of the misalignment scenario is the presence of isocurvature fluctuations in the dark matter distribution, due to quantum fluctuations of the axion field during inflation.
The quantum fluctuations generated over the $\NCMB\simeq 8$ $e$-folds that we can observe in the CMB fluctuations have a typical size of order $\HI \sqrt{\NCMB}/(2\pi)$.
When translated into spatial fluctuations of the misalignment angle $\theta_i=a_i/f_a$, the suppression factor $\HI/f_a$ makes these fluctuations tiny over most of the parameter space that we considered in Fig.~\ref{fig:AxionDM-abundance}. 
They can be observable, however, when the displacement from $\pi$ is very small, where even small fluctuations can be important.
We estimate the typical fluctuations $\Ddth$ of the displacement $\dth$ of the axion field from its minimum during inflation over a time of $\NCMB$ $e$-folds as $\Ddth \sim (\HI \sqrt{\NCMB})/(2\pi f_a)$. 
The corresponding isocurvature fluctuations in the energy density are \cite{Arvanitaki:2019rax}
\begin{equation}
\dIso \equiv \frac{\delta \rho}{\rho} \simeq C \frac{\Ddth}{\dth\, \ln \dth} \,,
\end{equation}
where $C\simeq 1.5-2.5$ is a constant with mild dependence on $\dth$.
The power spectrum of isocurvature density fluctuations is currently bound by the Planck experiment to be below $\dIso^\text{Planck} = 0.038 A_s$, where $A_s$ is the measured amplitude of the power spectrum of adiabatic fluctuations \cite{Akrami:2018odb}.
We correspondingly mark in orange the excluded regions along the white lines in Fig.~\ref{fig:AxionDM-abundance}.
A future detection of isocurvature fluctuations in the matter distribution would be a hint in favour of axion DM generated through the ``large-misalignment mechanism''. 
The expected reach on isocurvature fluctuations for CMB Stage 4 experiments is down to $\dIso^\text{CMB S4} =0.008 A_s$, improving Planck by a factor of 5 \cite{Abazajian:2019eic}.
We mark in yellow in Fig.~\ref{fig:AxionDM-abundance} this window of opportunity.

\item \textbf{Smaller decay constant and heavier axion.} 
Large anharmonic effects cause the delay of axion oscillations which, in turn, imply an enhancement of its relic density. 
This enhancement results in a lower decay constant to reproduce the observed DM abundance $\ODM h^2=0.12$. 
As shown before (see Eq.~\ref{initial_misalignment}) in our framework the initial condition for the axion field after inflation, $\theta_i=\pi-\delta\theta$, is dynamically determined by fundamental parameters of the theory. 
This allows us to estimate the axion relic abundance as a function of the axion decay constant and the Hubble parameter during inflation: $\Omega_a=\Omega_a(f_a,\HI)$%
\footnote{Under the reasonable assumption that the QCD scale does not change during inflation. The case $\Lambda_\text{QCD}^{\text{inf}}\neq\Lambda_\text{QCD}$ is straightforwardly obtained by changing the axion mass during inflation accordingly.}.
In Fig.~\ref{fig:AxionDM-abundance} we plot the axion mass as a function of $\HI$ for two different relic abundances: $\Omega_a =\ODM$ and $\Omega_a=0.1\ODM$. 
The iso-misalignment lines, corresponding to a constant $\delta\theta$ are also shown. 
It is worth commenting that the most interesting region, where the axion starts to oscillate very close to the top of the potential, that is, $\delta\theta\leq 10^{-8}$, corresponds to an approximately constant axion mass of $m_a\sim O(1\,{\rm meV})$.

A meV mass (THz frequency) QCD axion is currently beyond the reach of existing experimental searches for axion dark matter~\cite{Budker:2013hfa,Arvanitaki:2014dfa,Sikivie:1983ip,Dreyling-Eschweiler:2014mxa,TheMADMAXWorkingGroup:2016hpc,Andriamonje:2007ew,Anastassopoulos:2017ftl,Asztalos:2009yp,Armengaud:2014gea,Majorovits:2016yvk,Kahn:2016aff,Arvanitaki:2017nhi,Baryakhtar:2018doz}. While there are well established precision techniques to look for optical photons~\cite{Irwin2005,Rosfjord:06,GaoMazin,Mazin,Lita:08} and radio photons~\cite{Brubaker:2016ktl,TheMADMAXWorkingGroup:2016hpc}, the frequency range between $0.1$ to $10$ meV is the most challenging. Recently, many new experimental devices are considered as single photon sensors in the THz frequency range~\cite{Downes:2019nvi,Karasik:2012rb,Fink:2020noh}. However, it remains to be seen whether some of these techniques can be improved to have sufficiently good efficiency and very low background level to be useful for axion dark matter searches.
\end{enumerate}

\subsection{Model dependent constraints}
As shown in Fig.~\ref{fig:AxionDM-abundance}, our mechanisms require a low Hubble scale during inflation of order an eV. 
Therefore, the maximal temperature after inflation, is also bounded from above (see Eq.~\ref{RH_temp}): 
\begin{equation}
\TRH\sim 30\text{ TeV}\left(\frac{\HI}{\text{1 eV}}\right)^{1/2}\,.
\end{equation}
However, to thermally restore the symmetry after reheating, that is, to have a positive curvature at the origin of the potential, we need to satisfy the condition:
\begin{equation}
\TRH\gg \langle{\phi}\rangle\sim m_\phi\sim m_q\,.
\end{equation}
This means that the new vector-like quarks $q$ and real scalar $\phi$ are potentially accessible in collider experiments. Experimental searches for these new states can provide complementary information on the parameters of the theory. 
In the following we briefly summarise these constraints.

\subsubsection{Mixing with SM quarks and CKM unitarity}
\label{sec: mixing SM q}
Depending on its quantum numbers the new coloured fermion can mix with SM quarks. For example, we can have the vector-like quark $q$ with quantum numbers
\begin{equation}
q_{L,R}\sim(\mathbf{3},\mathbf{1},-1/3)\,,
\end{equation}
such that it mixes with down-type quarks (the case of an up-type vector-like quark would be analogous).
The relevant Yukawa couplings are given by
\begin{equation}\label{Yuk_Lag}
\mathcal{L}_q \supset y\phi\overline{q} q+ 
\Big( y_i\overline{q}_L\phi d^i_R+\widehat{y}_{i}\overline{Q}_L^iH q_R+\text{h.~c.} \Big)
\end{equation}
The first term corresponds to the Yukawa coupling of the new vector-like quark with the real scalar field, while the last terms are a mixing of the vector-like quark with the right-handed (RH) and left-handed (LH) components of the down-type SM quarks. 
The Lagrangian in Eq.~\eqref{Yuk_Lag} is the most general Lagrangian that respects the $\mathbb Z_2$ symmetry defined after Eq.~\eqref{eq: Yukawa phi q}.
After electroweak symmetry breaking, this leads to the $4\times 4$ quark mass matrix
\begin{equation}
M_q=\left(\begin{array}{cccc}
 & & &\widehat{y}_1v_\textsc{ew} \\
 & m^d_{3\times 3} & & \widehat{y}_2v_\textsc{ew}\\
  & & &\widehat{y}_3v_\textsc{ew}\\
  y_1\langle\phi\rangle& y_2\langle\phi\rangle&y_3\langle\phi\rangle &y\langle\phi\rangle\\
\end{array}\right)\,,
\end{equation}
where $m^d_{3\times 3}$ is the SM mass matrix for the down-type quarks.
The new Yukawa couplings, $y_i$ and $\widehat{y}_j$, can be complex in general but this does not affect the mechanism proposed in this work. The reason is that the relevant quantity entering in the theta term (see Eq.~\ref{theta}) is 
\begin{equation}
\arg\big[\det M_q \big]\,.
\end{equation} 
Thanks to the properties of determinants, this quantity changes exactly by $\pi$ as the scalar field flips sign, $\langle\phi\rangle\rightarrow-\langle\phi\rangle$, regardless of its value during inflation.\footnote{Note that even in the case the absolute value of the VEV is different before and after inflation, the change in $\arg[\det M_q]$ is exactly $\pi$.}

Nevertheless, the sizes of the new Yukawa couplings are indirectly constrained due to the mixing with the LH component of the SM quarks. The mass matrix above is diagonalised by a bi-unitary transformation
\begin{equation}
M_\mathrm{diag}=U_LM_qU_R^\dagger\,,
\end{equation}
where $U_{L/R}$ are unitary matrices relating the gauge and mass eigenstates.
The mixing angles describing the mixing of the vector-like quark with the LH part of the SM quarks is roughly given by
\begin{equation}
\theta_{i q}\approx\frac{|\widehat{y}_i|v_\textsc{ew}}{y\langle\phi\rangle}\,\,\,\text{   with  }i=1,2,3\,.
\end{equation}
The CKM matrix, $V_\textsc{ckm}=U_L^uU_L^{d \dagger}$, receives small corrections due to the aforementioned mixing of the LH components. 
One could be worried about the possibility of non-unitarity in the $3\times 3$ SM block induced by these new mixing angles. 
However, the angles above are generically small when the vector-like quark is heavier than a TeV, as it is our case. For example, $\widehat{y}_i\sim \mathcal O(0.1)$ gives a mixing angle
\begin{equation}
\theta_{i q}\leq 10^{-2}-10^{-3}\,,
\end{equation}
for $y\langle\phi\rangle\sim \mathcal O(1)$ TeV. Therefore the departures from CKM unitarity, expected to be of order $\theta_{i q}^2$, are tipically very small. 
We refer the reader to \cite{Aguilar-Saavedra:2013qpa} for a more detailed analysis and other flavour constraints of vector-like quarks.

The real scalar field can, in principle, mix with the SM Higgs. 
This mixing arises mainly from cubic and quartic terms in the scalar potential
\begin{equation}
V_\text{mix}=\mu_\text{mix}\phi (H^\dagger H)+\xi \phi^2(H^\dagger H)\,,
\end{equation}
where $H\sim (\textbf{1},\textbf{2},1/2)$ is the Higgs doublet. These terms generate a non-zero mixing of the SM Higgs and $\phi$, which is proportional to 
\begin{equation}
\sin \alpha\approx \frac{\mu_\text{mix}m_H}{m_\phi^2}+\xi\frac{m_H^2}{m_\phi^2}\,,
\end{equation}
and is expected to be small, since $m_H\ll m_\phi$. 
In addition we note that the small mixing limit ($\xi,\,\, \mu_\text{mix}/m_\phi\ll 1$) is approximately stable against quantum corrections. 
The reason is that these corrections are proportional to themselves or are suppressed by two powers of the small mixing angle between the vector-like quark and the SM quarks. 
For this reason we expect that the impact of the Higgs mixing on the phase transition of $\phi$ is negligible.

\subsubsection{Collider bounds}
There are different searches for scalar singlets and new vector-like coloured fermions at colliders. The coloured vector-like quarks $q$ can be both pair produced and singly produced at the LHC.
ATLAS and CMS collaborations have specific searches to look for its decay. 
These decays are expected to occur mainly in the channels:
\begin{equation}
\begin{split}
q\rightarrow bW\,,\,\,\, q\rightarrow tH\,,\,\,\, q\rightarrow tZ\,,\\
q\rightarrow tW\,,\,\,\, q\rightarrow bH\,,\,\,\, q\rightarrow bZ\,,
\end{split}
\end{equation} 
depending on whether the vector-like quark mixes with up-type or down-type quarks (see \cite{Aguilar-Saavedra:2013qpa} for details). 
The ATLAS and CMS collaborations look for these exotic coloured fermions in different final states with \cite{Sirunyan:2018qau,Aaboud:2018saj} or without \cite{Sirunyan:2017pks} lepton pairs. 
The decay $q\rightarrow bW$ excludes the vector-like quark mass $m_q\leq 1295$ GeV at $95\%$ C.L.~\cite{Sirunyan:2017pks}. 
If the dominant decay is $q\rightarrow qZ$, however, the limit is $m_q\geq 1280$ GeV for up-type vector-like quarks and $m_q\geq 1130$ GeV for down-type vector-like quarks \cite{Sirunyan:2018qau}. 
For intermediate cases, with branching ratios
\begin{equation}
	\text{Br}(q\rightarrow qW)\sim 
	\text{Br}(q\rightarrow q^\prime H)\sim 
	\text{Br}(q\rightarrow q^\prime Z)< 1\,,
\end{equation}
the constraint is milder and around 1 TeV (see \cite{Sirunyan:2018qau,Aaboud:2018saj,Sirunyan:2017pks} for details). To conclude, we will take $m_q\geq 1.3$ TeV as the standard lower bound of the vector-like quark masses.

\subsubsection{Cosmological considerations}
The requirement of reheating the new sector, $\TRH\gg \langle{\phi}\rangle$, implies the thermal production of coloured fermion and scalar fields in the early Universe. The scalar $\phi$ can decay into SM Higgses or coloured fermions, but the coloured fermion could be stable. This is because depending on the hypercharge, there may exist a conserved quantum number $U(1)_q$. In this case, when the temperature of the Universe drops below the QCD scale and confinement occurs, the heavy vector-like quarks $q$ can hadronise together with SM quarks to form super-heavy hadronic states. The presence of these neutral or electrically charged stable, super-heavy hadronic matter is severely constrained by a variety of searches \cite{Perl:2009zz,Burdin:2014xma}.  

However, the vector-like quark $q$, generically, is unstable due to mixing with SM quarks, which breaks $U(1)_q$ explicitly:
\begin{equation}\label{mixing}
\widehat{y}_i\overline{Q}^i_L H q_R\,,\quad y_j\overline{q}_L\phi d^j_R\,\,\,.
\end{equation}
Here we have assumed the vector-like quark $q$ is a $SU(2)_L$ singlet with hypercharge $-1/3$ (see also Eq. \ref{Yuk_Lag}). 
It is important to notice that the same decay that makes the fermion cosmologically harmless is, in fact, the same decay channel that is used in collider searches to look for it. 

When the coloured fermion, $q$, decays it can also be constrained by BBN. 
The decay of a neutral or charged massive particle, $q$ in this case, can modify dramatically  the abundance of light elements \cite{Kawasaki:2004qu,Pospelov:2006sc,Jedamzik:2007qk}, spoiling the success of standard nucleosynthesis. 
Additionally, if the fermion is sufficiently stable to decay during or after thermalization era it can be constrained from CMB. This is because the energy injection of the decaying particle will generate spectral distortions that are severely constrained \cite{Hu:1993gc,Chluba:2011hw}. 
However, as long as the Yukawa couplings in Eqs.~\eqref{Yuk_Lag}, \eqref{mixing} are not unreasonably small, BBN and CMB constraints will not constrain the parameter space of our framework.

\section{Discussion}
\label{sec: remarks}
In this section, we would like to further comment on the measure problem. 
The measure problem, practically, pertains to the inability to compute the statistical probability to live in each individual minimum in an eternally inflating universe. 
Specifically, a universe in a higher energy vacuum with a larger Hubble scale inflates faster, however, this universe also has an increased probability of decaying into a lower energy vacuum state, which has a smaller Hubble expansion rate. 
Depending on the choice of measure one assigns, we could find ourselves to be more or less likely to exist in a universe with a larger or smaller Hubble rate. 
The measure problem has cast doubts on the possibility of taking advantage of inflationary dynamics in the landscape, as it diminishes predictability. 
This has made anthropic reasoning the cornerstone of much inflationary model building.

In this paper, we have raised a different possibility. 
In the models we have discussed, we can ensure that regardless of the solution to the measure problem, there is always a binary choice of the sign of a parameter or a combination of parameters that decides the preferred minimum during inflation. 
Then, provided we have a deterministic evolution for the field $\phi$ after inflation --- in turn depending on a different combination of parameters --- we can always choose to switch the minimum where the field lives between inflation and today.
In this way, we \textit{circumvent} the measure problem, that is, the inability to assign probabilities, and instead find a situation where probabilities can be successfully computed.
For example, in the model in Sec.~\ref{sec: model} this probability is around $20\%$. 
Throughout this paper, we have treated the {\it measure} as providing a preferred vacuum during inflation for the scalar field, corresponding to the local minimum of either larger or smaller energy density. 
In principle, the correct measure could provide a more complicated relation depending on the parameters of the scalar potential during inflation and the inflationary Hubble scale. 
Although we are not aware of such an example in the literature, it is not inconceivable that the correct measure could directly depend not only on the shape of the full scalar potential, but also e.g.~on the Yukawa coupling of the scalar to fermions.\footnote{We thank Anson Hook for raising this relevant point.}
Even if such a measure existed, it is rather unlikely that it would select a minimum during inflation that always coincides with the minimum chosen by the field during the thermal evolution as outlined in Sec.~\ref{sec: model}.

In some sense, this is a third example, following \cite{Hook:2019pbh} and \cite{Hook:2019zxa}, where we can search for signatures related to the presence of other vacua in the landscape, albeit indirectly. 
In a recent example some of the authors worked on \cite{Hook:2019zxa}, a {\it deterministic} Higgstory is achieved for large Hubble rates because the electroweak minimum disappears and only the high energy minimum of the Higgs persists during inflation. 
In that case, the full story of changing vacua is deterministic very much independently of the choices of parameters. 
Both the examples of this paper and of \cite{Hook:2019zxa} feature a period of reheating when the temperature is high enough to restore the relevant symmetries and erase almost all the information about the inflationary choice of minimum, which depends on the solution to the measure problem.
After this phase the field evolves into the minimum determined by the parameters of the theory.
The key point of these examples is that a signature of this change of vacuum is still stored in some sector that couples during inflation to the field that changes vacuum, but decouples (at least) during the thermal phase when the symmetries are restored and the relevant field changes its vacuum. 
In \cite{Hook:2019zxa}, such a sector contains the frozen, long wavelength inflationary perturbations. 
In the case presented here, this sector is the QCD axion, whose dynamics are frozen throughout the early universe until QCD phase transition.
Interesting signatures may be found in similar decoupled sectors, for example, various inflationary perturbations, gravitational waves, axions and other non-thermal dark matter candidates, and possibly many others that are beyond the imagination of the authors.

\acknowledgments
The authors thank Daniel Egaña-Ugrinović and José Ramón Espinosa for many useful discussions, and Asimina Arvanitaki, Anson Hook, Gustavo Marques-Tavares and Ken Van Tilburg for valuable discussions and comments on the draft.
The authors also thank
Prateek Agrawal,
Andrea Caputo, 
Savas Dimopoulos, 
Matthew Johnson,  
Zhen Liu and
Veronica Sanz
for useful conversations.\\
The authors acknowledge the KITP for its hospitality during the inception of this project and National Science Foundation under Grant No. NSF PHY-1748958. 
M.~R.\ acknowledges Perimeter Institute for hospitality during the early phases of this project and is supported by the grants FPA2017-85216-P (AEI/FEDER,  UE), PROMETEO/2018/165 (Generalitat  Valenciana), the  Red Consolider MultiDark FPA2017-90566-REDC and FPU grant FPU16/01907.\\
Research at Perimeter Institute is supported in part by the Government of Canada through the Department of Innovation, Science and Economic Development Canada and by the Province of Ontario through the Ministry of Colleges and Universities.

\appendix

\section{One-loop finite-temperature corrections}
\label{app: V finite T}
At finite temperature, the classical (tree-level) potential receives both quantum and thermal corrections, which are contained in the effective potential $V_{\text{eff}}(\phi,T)$. In this appendix, we will evaluate both of these corrections at the one-loop level, and show that they can be expressed in a simple form around the temperature of the phase transition. Derivations of the general results for finite and zero-temperature 1-loop corrections can be found in \cite{Quiros:1999jp}.

Starting from zero temperature, the effective potential receives quantum (Coleman-Weinberg) corrections. Using dimensional regularisation and the $\overline{\text{MS}}$ renormalisation scheme, the zero-temperature one-loop correction is given by
\begin{equation}
    V_1^\text{CW}=\frac{1}{64\pi^2}\bigg( m_\phi^4(\phi^2)\Big[\log\bigg({\frac{m_\phi^2(\phi)}{\mu_R^2}}\bigg)-\frac{3}{2}\Big]-12 m_q^4(\phi^2)\Big[\log\bigg({\frac{m_q^2(\phi)}{\mu_R^2}}\bigg)-\frac{3}{2}\Big]\bigg).
\label{eq: V_CW}
\end{equation}
Here $\mu_R$ is the renormalisation scale and $m_i^2(\phi)$ are the  field dependent masses squared for the scalar and fermion
\begin{equation}
\begin{split}
m_{\phi}^2(\phi) &= -\mu^2+\kappa_3\phi + \frac{\lambda}{2}\phi^2 \\
m_q^2(\phi) &= (y\phi)^2.
\end{split}
\label{eq:m2phi}
\end{equation}
All couplings are taken to be effective couplings defined at the renormalisation scale $\mu_R$. 

\begin{figure}[h!] \centering
\includegraphics[width=\textwidth]{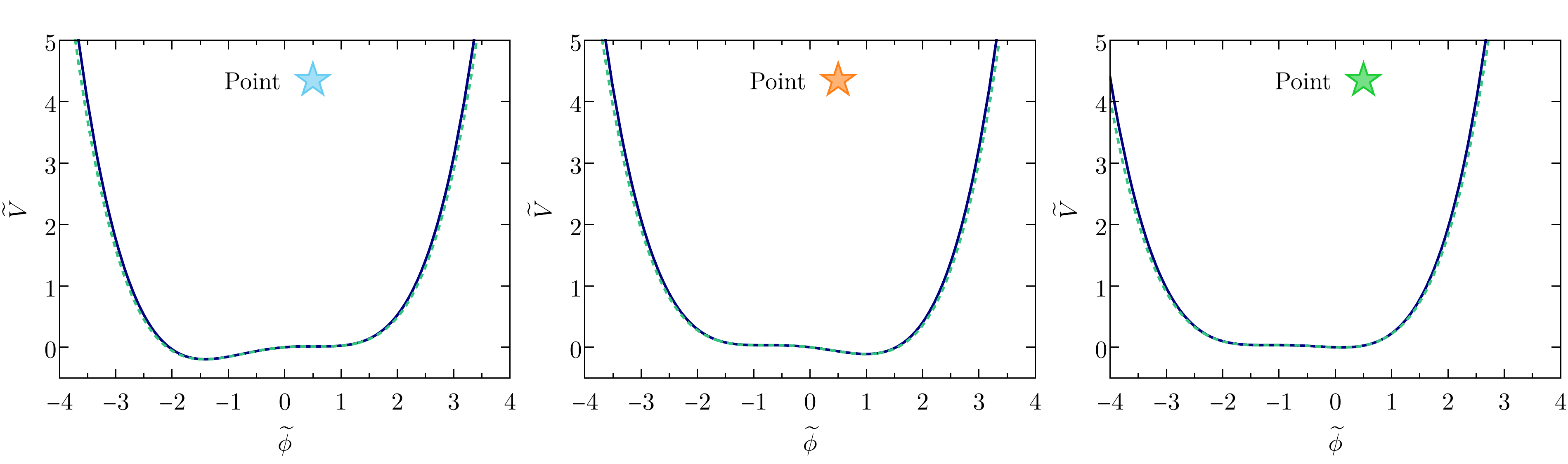}
\vspace*{-2em}
\caption{Comparison of the sum of the zero-temperature potential with either the full thermal \eqref{eq:finiteTcorrections} plus Coleman-Weinberg potentials \eqref{eq: V_CW} (light blue dashed line), or the leading terms in the high-temperature expansion \eqref{eq:Vrescaled} (solid blue line). 
The plots refer to the three points denoted with a star in Fig.~\ref{fig:MoneyPlot} (for which $\phi$ settles respectively in true vacuum, false vacuum and true vacuum after thermal decay) at the temperature $\Tt=\Tcat$, with $\yt=0.8$ and $\lambda=0.5$.}
\label{fig:compare Vth}
\end{figure}

The thermal one-loop correction to the effective potential is given by
\begin{equation}
V_1^T(\phi)=\frac{T^4}{2 \pi^2}\Big(\mathcal{J}_B\big[m_\phi^2(\phi)/T^2\big]  -12\mathcal{J}_F\big[m_q^2(\phi)/T^2\big]\Big).
\label{eq:finiteTcorrections}
\end{equation}
where $\mathcal{J}_i$ represent the thermal bosonic and fermionic functions, given respectively by
\begin{equation}
    \mathcal{J}_B\big[m^2(\phi)/T^2\big] = \int_0^\infty \mathrm dx\, x^2\, \log{\Big[1-e^{-\sqrt{x^2+m^2(\phi)/T^2}}\Big]}
\label{eq:scalarthermalfunction}
\end{equation}
\begin{equation}
    \mathcal{J}_F\big[m^2(\phi)/T^2\big] = \int_0^\infty\mathrm dx\, x^2\, \log{\Big[1+e^{-\sqrt{x^2+m^2(\phi)/T^2}}\Big]}, 
\label{eq:fermionthermalfunction}
\end{equation}
i.e.~they differ by a minus sign. The arguments of the thermal functions also depend on the field dependent masses (\ref{eq:m2phi}). Although one could evaluate these thermal functions numerically, we have taken an analytic approach in this paper in order to make the underlying physics more transparent.
A comparison between full and approximate result is given later in this Appendix in Fig.~\ref{fig:compare Vth} to justify this analytical treatment. 
We make use of the high-temperature expansions:
\begin{equation}
     \mathcal{J}_B\big[m^2/T^2\big] =  -\frac{\pi^4}{45}+\frac{\pi^2}{12}\frac{m^2}{T^2}-\frac{\pi}{6}\bigg(\frac{m^2}{T^2}\bigg)^{3/2}-\frac{1}{32}\frac{m^4}{T^4}\log{\frac{m^2}{a_bT^2}}
\label{eq:1loopbosonT}
\end{equation}
\begin{equation}
    \mathcal{J}_F\big[m^2/T^2\big] =  \frac{7\pi^4}{360}-\frac{\pi^2}{24}\frac{m^2}{T^2}-\frac{1}{32}\frac{m^4}{T^4}\log{\frac{m^2}{a_fT^2}}
\end{equation}
where $a_b=16\pi^2\exp{\big(3/2-2\gamma_E\big)}$ and $a_f=\pi^2\exp{(3/2-2\gamma_E)}$, and $\gamma_E$ is the Euler-Mascheroni constant.

The freedom to choose the renormalisation scale in the zero-temperature contribution to the effective potential leads to a very convenient cancellation of the quantum corrections with the logarithmic term in the high-temperature expansion of the thermal potential. 
As we are interested in the dynamics of the scalar field around the temperature of the phase transition, we choose $\mu_R \sim \Tca \sim \mu$. 
Consequently, the zero-temperature quantum corrections cancel out precisely at the phase transition, and we only need to consider the remaining thermal part. Furthermore, we also choose to neglect the remaining terms beyond leading order in the high-temperature expansion that did not cancel with the quantum corrections, as they are generally found to be subdominant around the VEV at the phase transition. Overall, we therefore keep only the tree-level potential and the leading order contributions from the high-temperature expansion of the thermal corrections:
\begin{equation}
    V_{\text{eff}}(\phi, T) =  \left(\kappa_1^3 +\frac{\kappa_3}{24}T^2\right) \phi+ \left(\frac{1}{24} T^2 \left(\lambda +12y^2\right)-\mu ^2\right)\frac{\phi ^2}{2}+\frac{\kappa_3  \phi^3}{3!}+ \frac{\lambda  \phi ^4}{4!}.
\label{eq:VeffLO}
\end{equation}
Rescaling to dimensionless quantities (see Table~\ref{tab: rescalings}) gives the potential of Eq.~\eqref{eq:Vrescaled} used in the main text.
The accuracy of this approximation compared to a numerical evaluation of the full one loop effective potential containing both zero and finite-temperature corrections can be seen in Fig.~\ref{fig:compare Vth}.
As visible from the plots, the high-temperature expansion is an excellent approximation for the range of parameters considered in Fig.~\ref{fig:MoneyPlot Yukawas}.
The expansion is close to the full result for the thermal corrections throughout the phase transition and until when the temperature decreases enough such that thermal corrections become negligible.
If the fermionic contribution is larger than the bosonic one, that is if $4N_c \yt^2 \gtrsim 1$ where $N_c=3$ is the number of colours, the approximation holds for $\sqrt \lambda \yt^2 \lesssim 1$.

Finally, it should be mentioned that the thermal perturbative expansion used thus far is known to become unreliable around the temperature of the phase transition \cite{Quiros:1999jp}. 
This is fixed through the procedure of daisy resummation of ring diagrams, which replaces the mass terms with effective masses, $m_\text{eff}^2=m^2+\Pi$. 
To leading order, the self-energy term $\Pi$ is proportional to $T^2$ but not to the field, and therefore also does not appear in the high-temperature expansion to leading order. For this reason, we do not consider daisy resummation in this paper.

\section[Phases of the explicitly broken $Z_2$ model]{Phases of the explicitly broken $\mathbb Z_2$ model}
\label{app: 3TC1}

For a scalar field potential which preserves $\mathbb Z_2$ symmetry, the phases in which there are either one or two minima of the potential, i.e. the unbroken and broken phases, are characterised by the sign of the coefficient of the $\phi^2$ term. In a model with explicit $\mathbb Z_2$ symmetry breaking, however, this classification depends on a combination of all the couplings in the potential. In this appendix, we will parametrise this dependence, and use it to determine two important results: the regions where the classical (zero temperature) potential contains either one or two minima, and, given that the classical potential contains two minima, the temperature when the second minimum first appears.

Although this phase classification will be seen to have a rather complicated dependence on the couplings in the potential, its derivation is fairly straightforward. 
To start, we remind the reader of the discriminant of a polynomial $\Delta$, a function of the polynomial's coefficients which can be used to either solve for or classify its roots. 
As we are interested in the roots of the derivative of the potential, we include here the explicit formula for the discriminant of a third degree polynomial, assuming all coefficients are real:
\begin{gather}
V'(\phi) = a +b\phi + c\phi^2 + d\phi^3 = 0 \,, \\ 
\Delta = b^2c^2 - 4b^3d - 4ac^3 -27a^2d^2 + 18abcd \,.
\end{gather}
The roots of a polynomial are all real and distinct provided that $\Delta>0$. 
With the further assumption that the coefficient $d$ (i.e.~the quartic coefficient in the potential) is positive, the potential will have two minima in the region $\Delta>0$. 
The solutions of $\Delta=0$ mark the transition between one and two minima. For the rescaled effective potential \eqref{eq:Vrescaled}, this condition is
\begin{multline}
\Delta = 1728 \left[6 \kat^3 \kct \left(\kct^2+3\right)+9 \kat^6-3 \kct^2-8\right]
 -432 \Tt^2 \left[12 \yt^2 \left(3 \kat^3 \kct-\kct^2-4\right)-\left(\kct^2+2\right)^2\right]\\
 -36 \Tt^4 \left[\kct^2+36 \left(\kct^2+8\right) \yt^4+24 \left(\kct^2+2\right) \yt^2+2\right] +\Tt^6 \left(12 \yt^2+1\right)^3
= 0 \,.
\label{3TC1condition}
\end{multline}
The first term, which is non-vanishing in the limit $\Tt\to 0$, describes the boundary of the region where the classical potential has one or two minima, given by the grey region in Fig.~\ref{fig:MoneyPlot}. 
At finite temperature, the equation $\Delta=0$ is a third degree polynomial in $\Tt^2$, and thus may have either one or three distinct, real positive solutions for $\Tt$, a feature in this model. 
The largest of these solutions, denoted $\Tt_0$ in the main text, describes the first appearance of the second minimum. 

\begin{figure}[h!] \centering
\includegraphics[width=.5\textwidth]{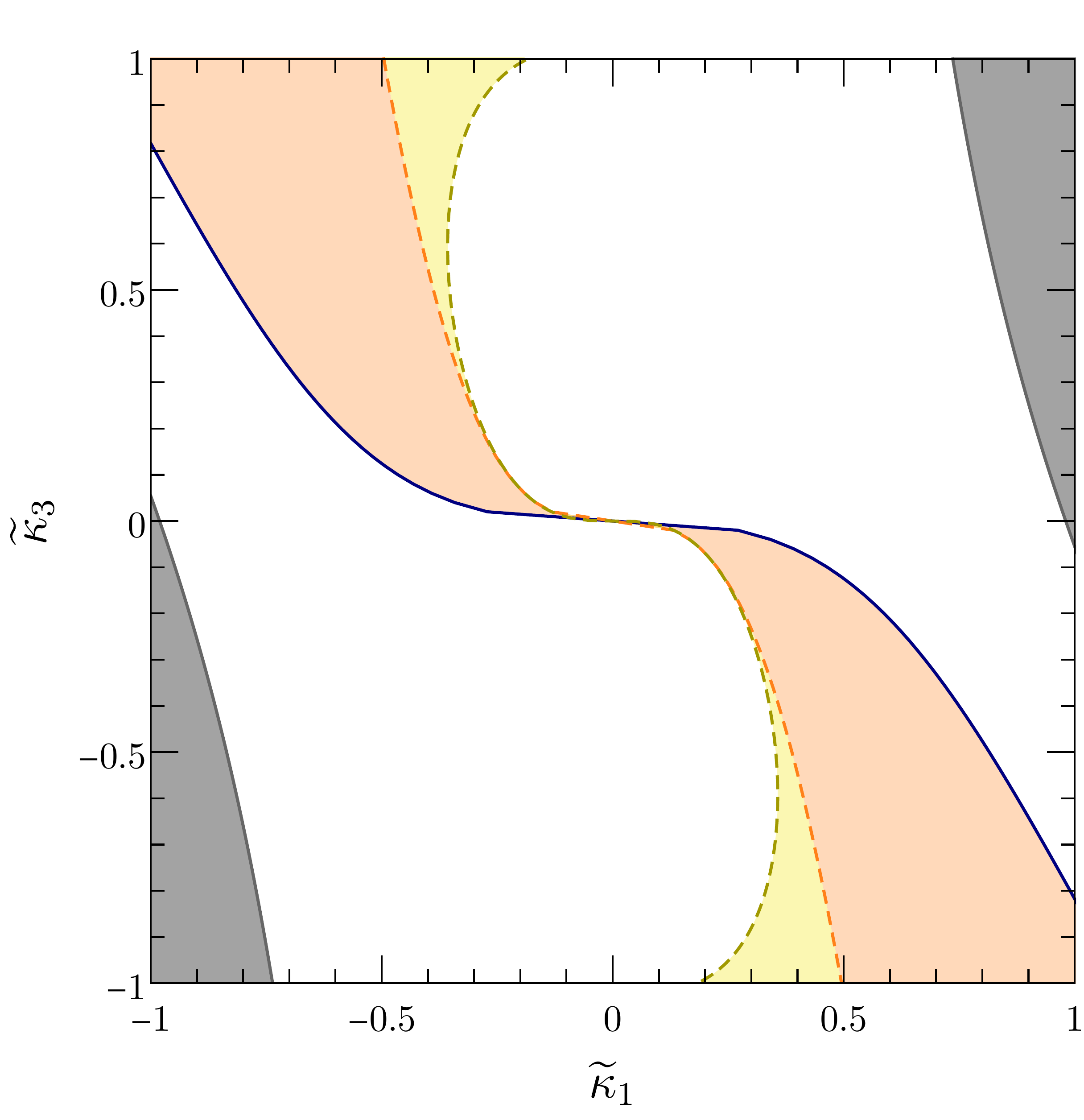}
\caption{Regions of parameter space where the scalar field $\phi$ settles directly in the true vacuum (in white), settles in the false vacuum after inflation before possibly thermally decaying (in orange), and where the thermal potential $V(\phi,T)$ incurs a two-stage rolling, first into the false then into the true vacuum, after inflation (in yellow) for $\yt=0.8$. 
The region shaded in grey corresponds to a single minimum potential.
The orange region is calculated using the condition given in Sec.~\ref{sec:vac selection}, whereas the yellow corresponds to the region where Eq.~\eqref{3TC1condition} has three real positive solutions for $\Tt$. 
The boundary of the yellow region corresponds to the line $\Tcat=\Tcbt$, where $\Tcbt$ is the temperature given by Eq.~\eqref{eq: Tc} where the two minima are degenerate,  and corresponds to the boundary of the region where the scalar field can roll into the false vacuum after inflation. The effects of vacuum decay are not included in this figure.}
\label{fig: 3 Tc1}
\end{figure}

What do these results mean for the evolution of the potential? 
As we have described in Sec.~\ref{sec: model}, the presence of explicit symmetry breaking effectively causes one of the minima of the classical potential to appear before the other as the universe cools. 
One may therefore assume that a simple criterion for the field to settle into the false  vacuum (up to the caveat of vacuum decay) is that the false minimum appears first. 
This analysis fails, however, in the region where there are three possible solutions for $\Tt$. 
In this case, the remaining two solutions correspond to the disappearance and later the reappearance of the false minimum. 
Here the field will first roll to the false minimum after inflation, but at a later stage when the false minimum disappears, it settles into the true minimum, which is always its final state. 
This new region, shaded in yellow in Fig.~\ref{fig: 3 Tc1}, borders the regions of true vacuum final state (in white) and the region where $\phi$ initially rolls to the false vacuum before possibly thermally decaying. 
Depending on the quartic coupling $\lambda$, as discussed in Sec.~\ref{subsec:Thermalvacuumdecay}, the part of the orange region bordering the yellow one can thermally decay, which is shaded in green in Fig.~\ref{fig:MoneyPlot}.
The criterion for vacuum selection described in Eq.~\eqref{deriv0}, corresponding to the orange dashed line in Fig.~\ref{fig: 3 Tc1}, correctly identifies the regions of a true and false vacuum final state.

\section{Vacuum decay}
\label{app: vacuum decay}

Here we detail the calculation of the vacuum decay rates at finite and zero temperatures using the TWA. The true vacuum bubble is an unstable solution to the Euclidean equations of motion \cite{Coleman:1977py}
\begin{equation}
(\partial_t^2+\nabla^2)\phi =\frac{\mathrm d V_\text{eff}}{\mathrm d\phi}\,,
\end{equation}
which under the assumption of spherical $O(d)$ symmetry take the form
\begin{equation}\label{Eq_bubble}
\left(\frac{\mathrm d^2}{\mathrm dr^2}+\frac{d-1}{r}\frac{\mathrm d}{\mathrm dr}\right)\phi =\frac{\mathrm d V_\text{eff}}{\mathrm d\phi}\,.
\end{equation}
In the TWA, the energy difference between the non-degenerate vacua, $\epsilon$, is small compared to the height of the barrier. When the size of the bubble is large compared to the wall thickness it is possible to neglect the second term in Eq.~\eqref{Eq_bubble} and it is reduced to \cite{Coleman:1977py,Linde:1981zj}:
\begin{equation}\label{thin_wall_1}
\frac{\mathrm d^2\phi}{\mathrm dr^2}=V_\text{eff}'(\phi)\,.
\end{equation} 
As we show below this allows a good analytical control of the zero-temperature and finite-temperature vacuum decay rates.

\subsection{Vacuum decay rate at zero temperature in the thin-wall approximation}
We can estimate the rate of false vacuum decay per unit volume as
\begin{equation}\label{zeroT_rate}
\Gamma/V=Ae^{-B}\,,
\end{equation}
where $B\equiv S_4$ is the Euclidean action for the bounce, which is $O(4)$ symmetric \cite{Coleman:1977py}. 
As shown by Coleman, after minimization, this action is given by:
\begin{equation}
S_4=\frac{27\pi^2S_1^4}{2\epsilon^3}\,.
\end{equation}
We use the shift $\phi\rightarrow \phi-\kappa_3/\lambda$ to get a potential where the only $\mathbb Z_2$ breaking part is a linear term:
\begin{equation}
V=V_{\mathbb Z_2}+V_{\not{\mathbb Z_2}}=\frac{\lambda}{4!}\left(v^2-\phi^2\right)^2+\left[\kappa_1^3+\frac{\kappa_3\left(\kappa_3^2+3\lambda\mu^2\right)}{3\lambda^2}\right]\phi\,.
\end{equation}
We only need to compute the one-dimensional action and the energy difference \footnote{Note that a factor 2 in the final expression of $S_1$ is missing in \cite{Coleman:1977py}.}:
\begin{equation}\label{surface_energy_bubble_zeroT}
S_1=\int^{v}_{{-v}} \mathrm d\phi\sqrt{2V_{Z_2}(\phi)}=\frac{2}{3}\sqrt{\frac{\lambda}{3}}v^3\,,
\end{equation}
\begin{equation}
\epsilon=2v(\kappa_1^3 + \frac{\kappa_3^3} {3\lambda^2} + \frac{\kappa_3 \mu^2}{
	\lambda} )\,, \text{  with VEV }v=\sqrt{3 (\kappa_3^2 + 2 \lambda \mu^2)/\lambda^2}\,.
\end{equation}
Finally, the semi-classical exponent is given by:
\begin{equation}
B\equiv S_4=\frac{27 \pi^2 S_1^4}{2 \epsilon^3}=\frac{81 \sqrt{3} \pi^2 (\kappa_3^2 + 2 \lambda \mu^2)^{9/2}}{\lambda (\kappa_3^3 + 3 \kappa_1^3 \lambda^2 + 
	3 \kappa_3 \lambda \mu^2)^3}=\frac{1}{\lambda}\frac{81\sqrt{3}\pi^2(\kct^3+2)^{9/2}}{ (3\kat^3+\kct^3+3\kct)^3}\,,
\end{equation}
where in the last step we have used the dimensionless units defined in Table~\ref{tab: rescalings}. 
The probability of false vacuum decay is obtained by multiplying the rate in Eq.~(\ref{zeroT_rate}) by the present space-time volume $H_0^{-4}$, where $H_0=1.5\cdot 10^{-33}\,\text{eV}$ is the Hubble parameter today.
To determine the rate we also need the prefactor, $A$, which by dimensional analysis arguments \cite{Linde:1981zj} can be estimated as:
\begin{equation}
A\sim\left(\frac{S_4}{2\pi}\right)^2\times \mu^4\,.
\end{equation}
As stated in Sec.~\ref{zero_T_decay}, the decay probability 
\begin{equation}
H_0^{-4}Ae^{-B}\,,
\end{equation} 
turns out to be much smaller than 1 in all the false vacuum region for $\lambda$ in the perturbative regime, rendering our false vacuum final state completely stable against zero-temperature decay.

\subsection{Vacuum decay rate at finite temperature in the thin-wall approximation}
In this subsection we compute $E_b(T)/T$ in Eq.~(\ref{thermal_rate}). 
Let us now consider the free-energy, or equivalently the three-dimensional action, of the bubble:
\begin{equation}\label{3_dim_action}
E_b(T)=S_3=4\pi\int^\infty_0 r^2 \mathrm dr\left[\frac{1}{2}\left(\frac{\mathrm d\phi}{\mathrm dr}\right)^2+V(\phi,T)\right]\,.
\end{equation}
The first term is, as before, a surface term which increases as $\sim r^2$ while the volume term goes as $\epsilon r^3$. 
Therefore, we get in the thin-wall approximation:
\begin{equation}
S_3=-\frac{4\pi}{3}r^3\epsilon+4\pi r^2S_1\,.
\end{equation}
Since the three-dimensional action of the bubble must be an extremum, we can determine the radius of the critical bubble as
\begin{equation}\label{semiclassical_exponent}
\frac{\mathrm dS_3}{\mathrm dr}=0\rightarrow r_c=\frac{2S_1(T)}{\epsilon}\rightarrow S_3(T)=\frac{16\pi S_1^3(T)}{3\epsilon(T)^2}\,.
\end{equation}
It is simple to show that this extremum is a maximum, which reflects the fact that the bubble is an $O(3)$ symmetric, unstable solution. 

We can perform the shift previously introduced in Eq.~\eqref{eq:amaliashift un-rescaled}, $\phi\rightarrow \phi-\kappa_3/\lambda$,  to compute the energy difference between true and false vacuum and write the potential as a $\mathbb Z_2$ symmetric plus a $\mathbb Z_2$-breaking term:
\begin{equation}\label{eq:Potential_Amalia_shift}
V=V_{\mathbb Z_2}+V_{\not{\mathbb Z_2}}=\frac{\lambda}{4!}\Big(\phi^2-v(T)^2\Big)^2+\left[\kappa_1^3+\frac{\kappa_3\left(8\kappa_3^2+24\lambda\mu^2-12T^2y^2\lambda\right)}{24\lambda^2}\right]\phi\,.
\end{equation}
The energy difference between true and false vacua is given by:
\begin{equation}\label{epsilon_T}
\epsilon(T)=\frac{y^2\kappa_3}{\lambda}\left[\Tcb^2-T^2\right]v(T)=\frac{\mu^3}{\sqrt{\lambda}}\yt^2\kct\left[\Tcbt^2-\Tt^2\right]v(T)\,,
\end{equation}
with $\Tcb$ and $v(T)$ given by:
\begin{equation}\label{Tc2 and VEV}
\begin{split}
&\Tcb=\sqrt{\frac{2}{3}}\frac{\sqrt{\kappa_3^3+3\kappa_1^3\lambda^2+3\kappa_3\lambda\mu^2}}{y\sqrt{\kappa_3}\sqrt{\lambda}}=\frac{\mu}{\sqrt{\lambda}}\frac{\sqrt{2}}{\yt}\sqrt{\frac{\kct^2}{3}+\frac{ \kat^3 }{\kct}+1}\,,\\&
v(T)^2=\frac{12\kappa_3^2-T^2\lambda(12y^2+\lambda)+24\lambda\mu^2}{4\lambda^2}=\frac{\mu^2}{\lambda}\left(6+3\kct^2-\frac{\Tt^2(1+12\yt^2)}{4}\right)\,.
\end{split}
\end{equation}
Here, again, we have used the dimensionless units defined in Table~\ref{tab: rescalings}.

Another temperature that we use to find the minimum exponent $E_{b}(T_\text{min})/T_\text{min}$ is $\Tst$. 
We define it as the temperature at which the curvature at the origin of the $\mathbb Z_2$-symmetric potential, $V_{\mathbb Z_2}$, (see Eq.~\eqref{eq:Potential_Amalia_shift}) flips sign. 
At $\Tst$ the thermal mass is zero and, therefore, $v(\Tst)=0$. This temperature is given by:
\begin{equation}\label{Tstar}
\Tst=\frac{2\sqrt{3}\sqrt{\kappa_3^2+2\lambda\mu^2}}{\sqrt{12y^2\lambda+\lambda^2}}=\frac{\mu}{\sqrt{\lambda}}\frac{2\sqrt{\kct^2+2}}{\sqrt{4\yt^2+1/3}}\,.
\end{equation}
Regarding the surface energy of the bubble, $S_1$, we can compute it as the action corresponding to the one-dimensional theory:
\begin{equation}\label{surface_energy_bubble}
S_1=\int^{v(T)}_{-v(T)}\mathrm d\phi\sqrt{2V(\phi,T)}=\frac{2\sqrt{\lambda}}{3\sqrt{3}}v(T)^3\,,
\end{equation}
computed in the limit $\epsilon\rightarrow 0$. 
With the quantities $S_1(T)$ and $v(T)$ above, we can go back to Eqs.~\eqref{3_dim_action}, \eqref{epsilon_T} and compute the semi-classical exponent $E_b(T)/T$ in the thin-wall approximation:
\begin{equation}\label{free_energy_appendix}
\frac{E_b(T)}{T}=\frac{128\pi}{243\sqrt{3}}\frac{ v(T)^9\lambda^{3/2}}{T\epsilon(T)^2}=\frac{1}{\sqrt{\lambda}}\frac{\pi}{27\sqrt{3}}\frac{(12\kct^2-\Tt^2(12\yt^2+1)+24)^{7/2}}{\Tt(6\kat^3+\kct(2\kct^2-3\Tt^2\yt^2+6))^2}\,.
\end{equation}
To check the validity of our results we need to ensure that we are in a regime in which the width of the wall, estimated as $\Delta r\sim (V''_\text{min})^{-1/2}$, is thin compared to the radius of the critical bubble:
\begin{equation}
r_c/\Delta r\gg 1\rightarrow\frac{2\sqrt{\lambda}S_1}{\sqrt{3}\epsilon}v(T)\gg 1\,.
\end{equation}
Finally, to determine whether the false vacuum is stable or not we find the temperature at which $B(T)=\frac{E_b(T)}{T}$ is a minimum. 
It turns out to be:
\begin{equation}
T_\text{min}=\frac{1}{2}\sqrt{(6 \Tcb^2 - 5 \Tst^2) + \sqrt{\left(6\Tcb^2-5\Tst^2\right)^2 + 8 \Tcb^2 \Tst^2}}\,.
\end{equation}
In Fig.~\ref{fig:MoneyPlot} we plot the stable region in the $(\kappa_1,\kappa_3)$ plane with the criterion of the minimum semi-classical exponent: if the minimum satisfies, $B(T_\text{min})>140$, we consider that region as stable (see Fig.~\ref{fig: thermal bounce}). 
It is unstable to thermally decay otherwise.


\providecommand{\href}[2]{#2}\begingroup\raggedright\endgroup

\end{document}